\documentclass[letter,12pt,bibliography=totocnumbered]{book}
\usepackage[utf8]{inputenc}
\usepackage{graphicx}
\usepackage{hyperref}
\usepackage{cleveref}
\usepackage{subfigure}
\usepackage{cite}
\usepackage{csquotes}

\usepackage[square,numbers,sectionbib,sort&compress]{natbib}

\makeatletter
\renewcommand\footnotesize{%
   \@setfontsize\footnotesize\@ixpt{10}%
   \abovedisplayskip 8\p@ \@plus2\p@ \@minus4\p@
   \abovedisplayshortskip \z@ \@plus\p@
   \belowdisplayshortskip 4\p@ \@plus2\p@ \@minus2\p@
   \def\@listi{\leftmargin\leftmargini
               \topsep 4\p@ \@plus2\p@ \@minus2\p@
               \parsep 2\p@ \@plus\p@ \@minus\p@
               \itemsep \parsep}%
   \belowdisplayskip \abovedisplayskip
}
\makeatother

\graphicspath{{Figures/}}

\begin{document}

\author{Leon Karpa}
\title{Trapping single ions and Coulomb crystals with light fields}
\date{}
\frontmatter

\maketitle

\chapter{Preface}

The scope of this book is on providing insight into the recently emerged field of optical trapping of ions. Ever since the ground-breaking introduction of light fields as tools for exerting trapping forces on matter in 1970 by Ashkin \cite{Ashkin1970}, optical dipole traps have been used with remarkable success in several fields of research, most notably in atomic physics, where they have enabled an unprecedented level of control over neutral atoms and molecules both at the level of quantum ensembles \cite{Bloch2008} as well as individual particles \cite{Endres2016,Barredo2018}.\\

Another tremendously influential development in Physics was enabled by the invention of radiofrequency traps pioneered by Paul \cite{Paul1990} and Dehmelt, allowing for confinement of single atomic ions demonstrated by Wineland \cite{Wineland2013} decades before this was realized on the level of individual neutral atoms in optical traps. Ions provide a number of properties that render them ideal for state-of-the-art applications in several cutting-edge areas of contemporary research ranging from metrology \cite{Rosenband2008}, quantum simulation and computation \cite{Blatt2008} to ultracold chemistry \cite{Cote2002}.\\

However, it was found recently that in some situations it is highly advantageous to confine ions without employing any radiofrequency-based Paul traps or strong external magnetic fields in Penning traps, e.g. when investigating the interaction of neutral atoms and ions in the regime of ultralow interaction energies \cite{Haerter2014}. Adapting optical traps for ions \cite{Schaetz2017} is a promising way to approach such scenarios and the focus of this work is to present a comprehensive overview of the background and concepts behind this technique as well as to discuss the currently achievable level of control, encountered limitations and perspectives for future applications.\\

\newpage
Chapter 1 is intended to provide a context of the described experiments within the field of atomic physics. It highlights the advantages of the currently used techniques applied for trapping and manipulating ions in comparison with the features granted by optical traps for neutral atoms. \\

Chapter 2 lays out the general concepts, requirements and methodology for realizations of optical trapping of ions.\\

Chapter 3 describes how the tools and methods presented in the previous chapter can be used to confine single atomic ions in optical traps. Individual developments achieved in the last years are highlighted by presenting and discussing milestone experiments \cite{Schneider2010,Enderlein2012,Huber2014,Lambrecht2017}. The aim of this chapter is to provide an understanding of the current performance granted by this approach and its prospective possibilities for applications in the near future. It concludes with a discussion of the identified limitations and strategies for future improvement.\\

Chapter 4 is focused on experiments reaching beyond the established optical traps for single ions. It provides an in-depth view on the recently demonstrated extension of the presented methods to the case of ion Coulomb crystals \cite{Schmidt2018}.\\

Chapter 5 highlights the key features of the concepts discussed in this book, summarizes the main features of optical traps for ions and provides an outlook for future applications building on the capabilities afforded by these novel tools.\\

\begin{flushright}
Leon Karpa
\end{flushright}

\chapter{Dedication}
This work is dedicated to my family.

\tableofcontents

\mainmatter

\chapter{Introduction} \label{chpt:Intro}

\begin{figure}[h!]
	\subfigure[]{
		\includegraphics[width=0.3\textwidth]{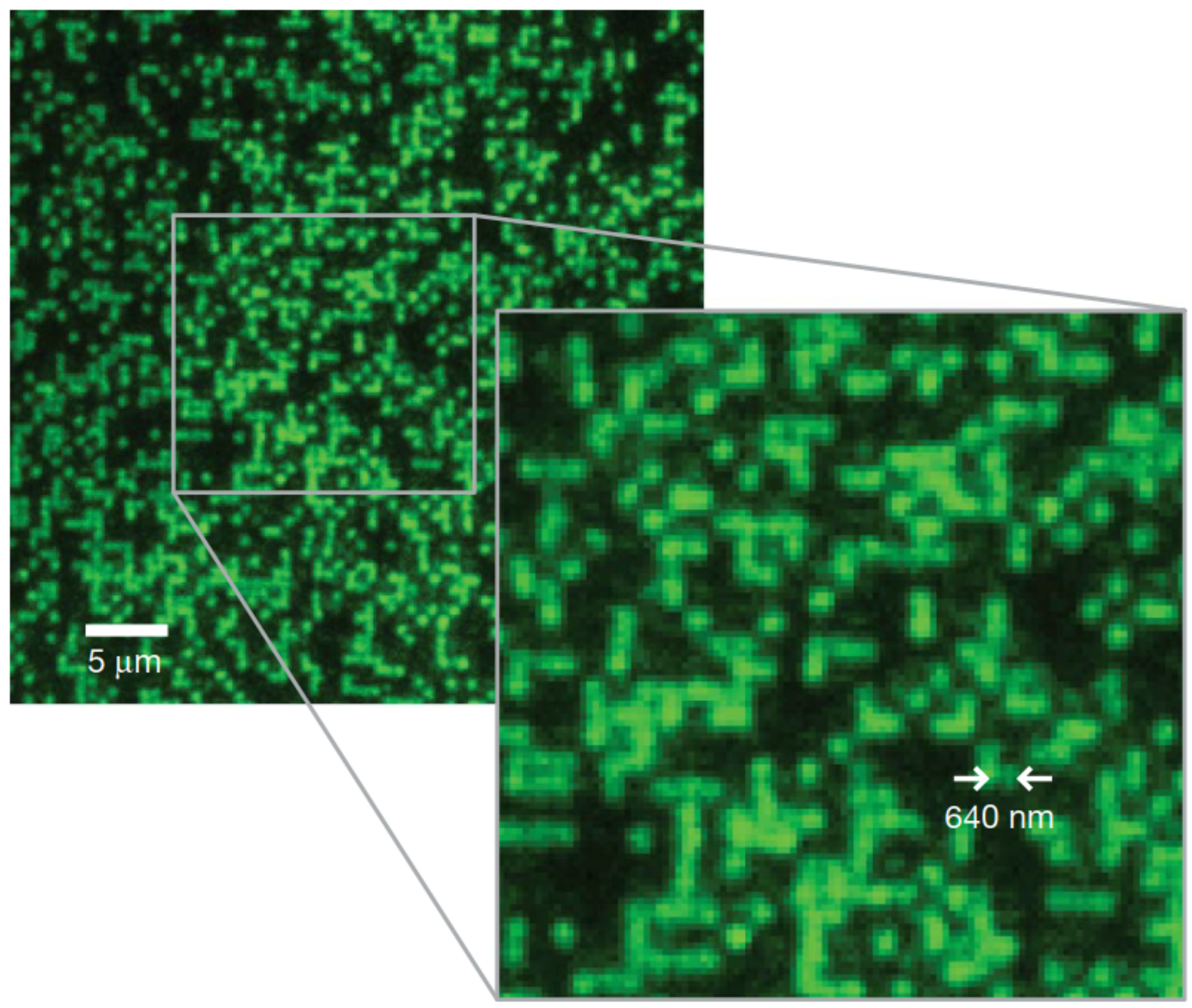}
		\label{fig:QuantGasMikroscope}
	}
	\subfigure[]{
		\includegraphics[width=0.7\textwidth]{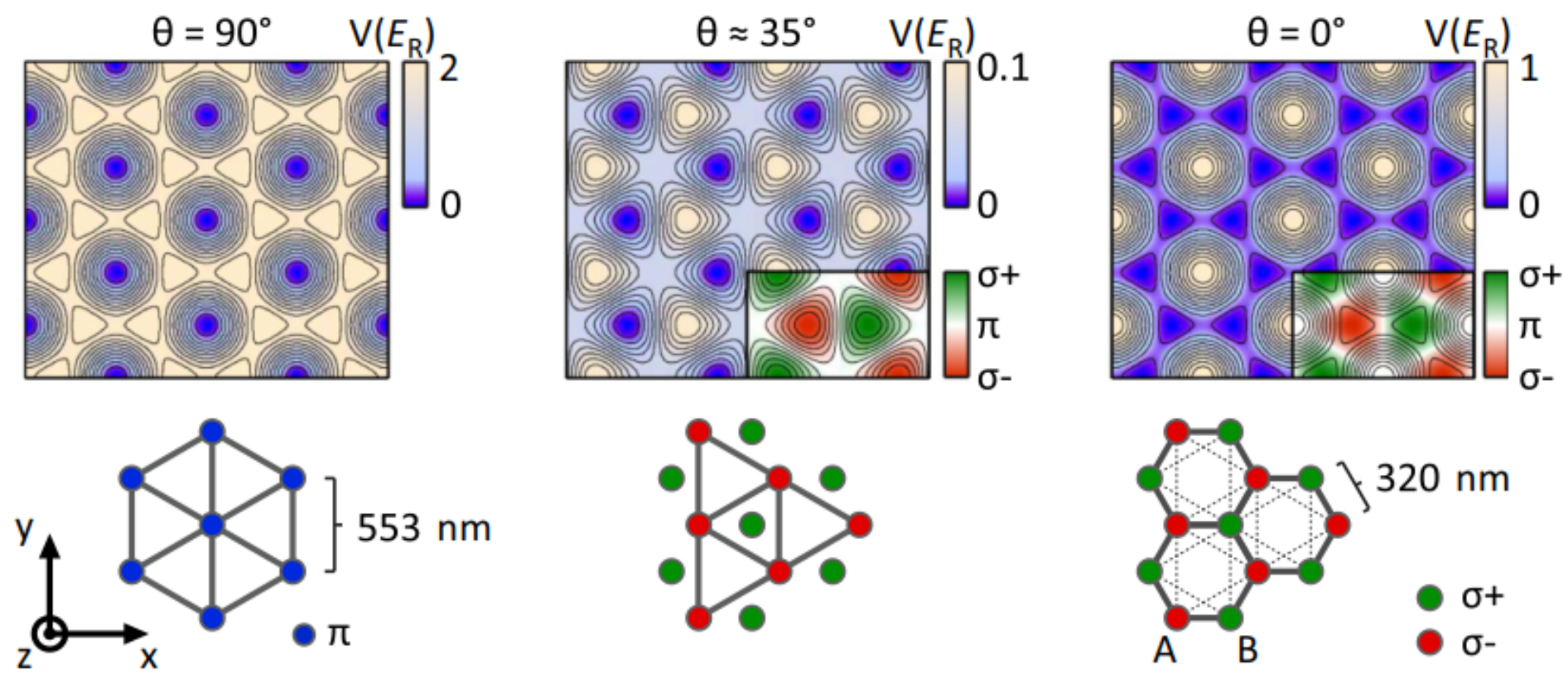}
		\label{fig:OptLattice}
	}
	\caption[]{Optical trapping potentials in experiments with neutral atoms: \textbf{(a)} individual  atoms imaged within the optical potential of a standing wave (courtesy of W. Bakr, taken from \cite{Bakr2009}) and \textbf{(b)} versatile optical trapping  geometries (from left to right: triangular, polarization, and honeycomb lattice) obtained by using a superposition of three laser beams and controlling their respective polarizations (courtesy of K. Sengstock, taken from \cite{Luehmann2014})\footnotemark[1].}
\end{figure}

Over the past decades, optical forces have proved to be a pivotal instrument for confinement and manipulation of atoms \cite{Phillips1998,Chu1998,Cohen-Tannoudji1998}. Their use has enabled highly versatile trap geometries while at the same time preserving the capability to investigate interactions of atomic gases and molecules at ultralow temperatures \cite{Bloch2008}. Optical traps are now routinely used to trap and control individual neutral atoms and atomic ensembles on the nanoscale with sub-wavelength resolution, e.g. in the Mott insulator regime \cite{Bloch2005} or in the demonstrated realizations of quantum gas microscopes \cite{Bakr2009,Sherson2010} (Fig. \ref{fig:QuantGasMikroscope}). Recently these techniques have been once more tremendously improved by refining the control over combinations of several optical fields and making use of the available degrees of freedom such as geometry, intensity, polarizations and relative phases \cite{Greif2015,Morsch2006,Bloch2008}. Consequently, this progress has led to the realization of superlattices \cite{Kangara2018}, and Kagom\'{e} lattices \cite{Jo2012} (Fig. \ref{fig:OptLattice}), as well as to the assembly of linear \cite{Endres2016}, two-dimensional \cite{Barredo2016} and even three-dimensional arrays \cite{Barredo2018} of atoms.\\

\makeatletter
\@addtoreset{footnote}{section}
\makeatother

For several decades, the exquisite degree of control over individual atoms has been one of most prominent characteristic features in the field of ion trapping, a vastly successful and rapidly expanding area of atomic physics \cite{Paul1990,Wineland2013}. In this field of research, individual atoms are typically confined in radiofrequency or Penning traps, affording preparation, operation and detection fidelities approaching unity and storage times on the order of months or years. 
Motional and internal electronic degrees of freedom of atomic ions can be manipulated on the quantum level with individual addressability and exceptionally high operation fidelities in excess of 99$\%$ \cite{Monroe2013,Gaebler2016,Ballance2016}, making trapped ions one of the most promising and performant platforms for quantum information processing (QIP), quantum simulations and metrology to date. Long-range interactions that are difficult to establish in experiments with neutral atoms allow for using phonon modes of Coulomb crystals as a bus for generating entanglement between ions in linear chains \cite{Blatt2012}.\\

\begin{figure}
	\subfigure[]{
	\includegraphics[height=0.3\textwidth]{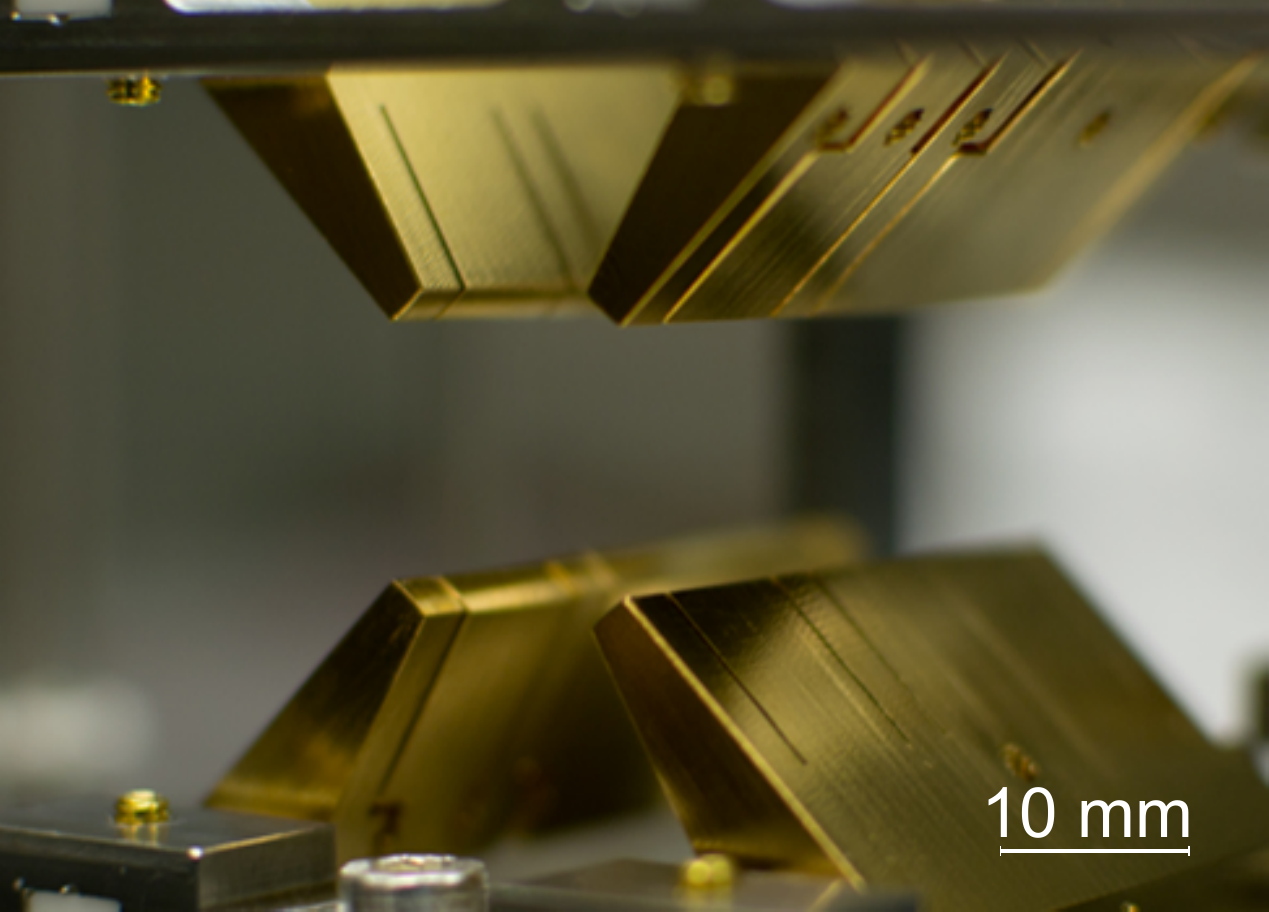}
				}
			\label{fig:PaulTrapPicture}
	\hspace{.05\textwidth}
	\subfigure[]{
	\includegraphics[height=0.3\textwidth]{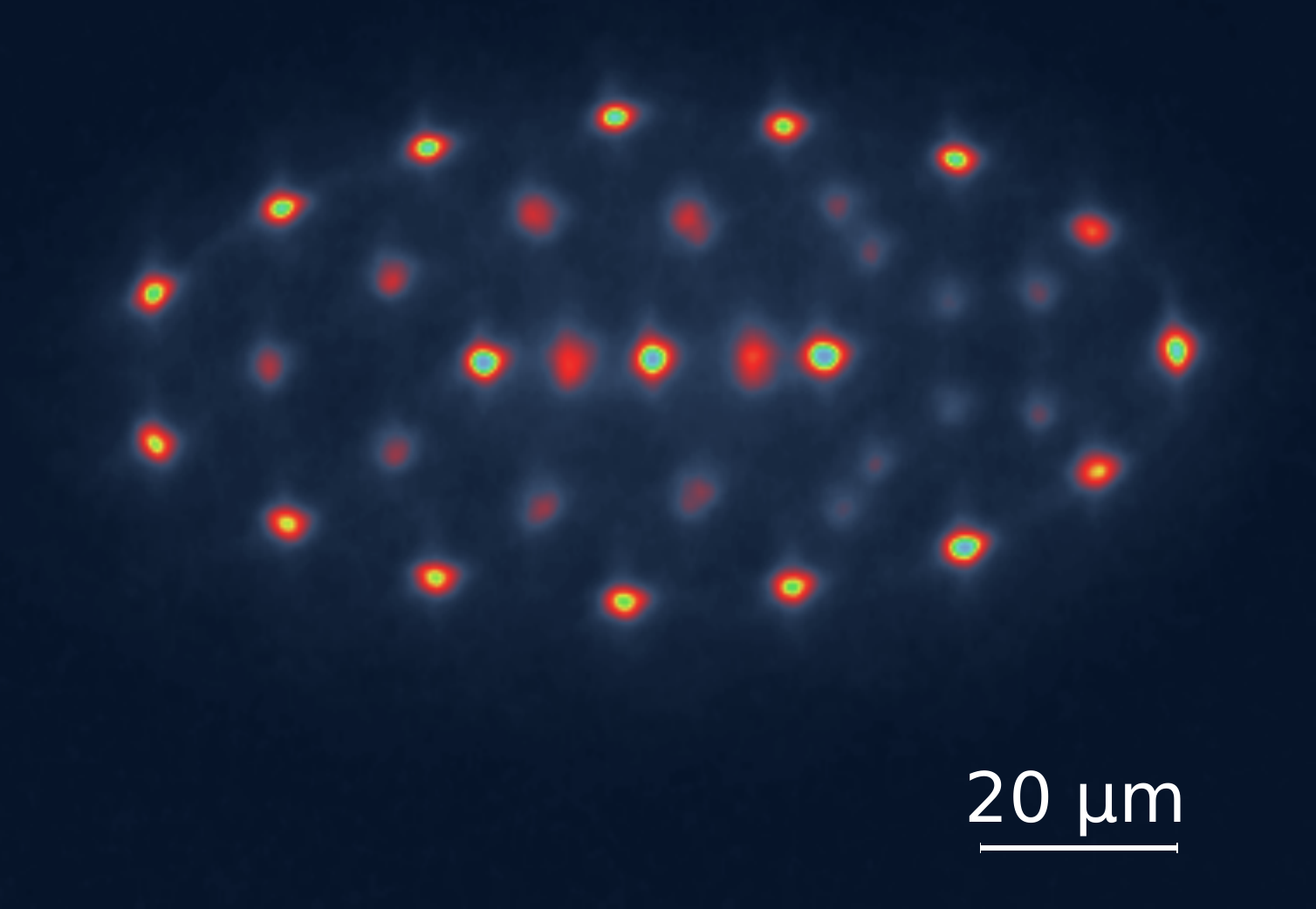}
				}
			\label{fig:IonCrystalPicture}
	\caption{\textbf{(a)} Linear Paul trap with segmented electrodes. The diagonal spacing between the electrodes is $ 18 \, \mathrm{mm} $. \textbf{(b)} False-color fluorescence image of a three-dimensional ion Coulomb crystal. The spacing between the ions is determined by the choice of the trap parameters. Typically, the ions are separated by approximately $ 20 \, \mu \mathrm{m} $ allowing for individual addressability.}
	\label{fig:PaulTrapAndIons}
\end{figure}

\footnotetext[1]{Reprinted figure with permission from D.-S. L\"{u}hmann, O. J\"{u}rgensen, M. Weinberg, J. Simonet,	P. Soltan-Panahi, and K. Sengstock. Physical Review A, 90:013614 (2014). Copyright 2019 by the American Physical Society.}

\begin{figure}
	\begin{center}
		\includegraphics[height=0.4\textwidth]{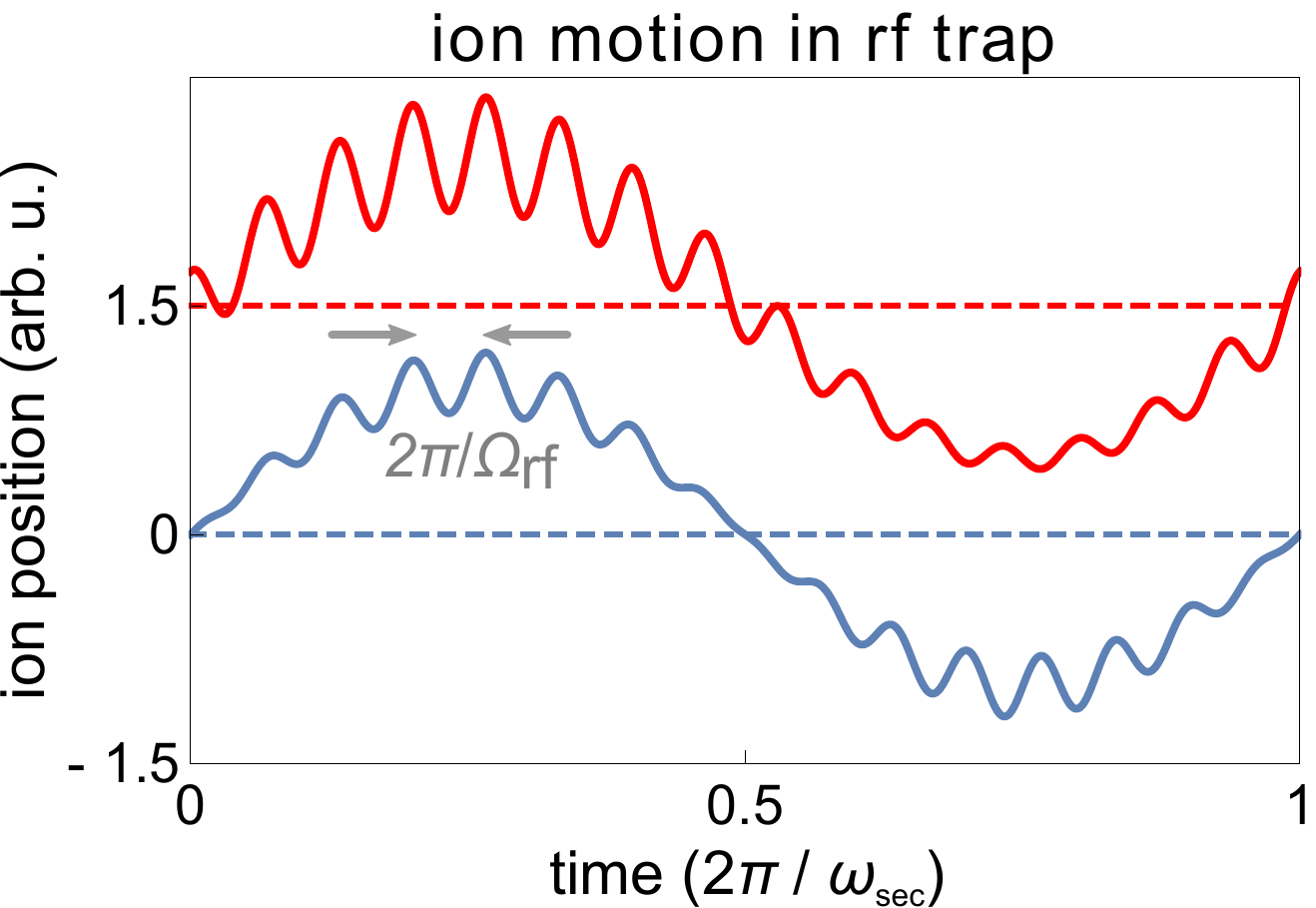}
	\end{center} 
\caption{Motion of an ion in a radiofrequency trap. The oscillation at a frequency $ \omega_{sec} $ in the quasi-harmonic trapping potential is superimposed with a rapid oscillation at the modulation frequency of the trap, $ \Omega_{rf}$. In the presence of any stray electric field the amplitude of this micromotion is non-zero at all times (upper trace).}
	\label{fig:MuMotion}	
\end{figure}

While Paul traps have proved to be very successful or leading platforms for applications in many fields of contemporary research, some outstanding challenges closely related to limitations in scalability have emerged. The generation of nanoscale potentials with Paul traps remains another challenge despite of rapid advances achieved with planar trapping geometries. An intrinsic property of such radiofrequency (rf) traps is the driven motion at the modulation frequency which is superimposed with the secular motion of ions as illustrated in figure \ref{fig:MuMotion}. While this so-called micromotion can be neglected in most applications based on the manipulation of common mode excitations within linear ion crystals, it can impose a fundamental limitation on other experiments sensitive to kinetic energy \cite{Cetina2012,Tomza2019}. A prominent example is the rapidly growing field of ion-atom interactions \cite{Smith2005,Grier2009,Schmid2010,Zipkes2010,Rellergert2011,Deiglmayr2012,Meir2016,Kleinbach2018,Fuerst2018,Dutta2017,Haze2018} where the presence of driven electromagnetic fields typically limits the accessible collision energies to the range of 1 mK \cite{Haerter2014,Tomza2019}. While some approaches exploiting the favorable kinematic properties of specific combinations of neutral atom and atomic ion species confined in conventional Paul traps exist \cite{Cetina2012,Fuerst2018} and seem promising for accessing a range of collision energies that could approach the quantum interaction regime, the generalization to either generic atom-ion combinations or to experiments involving more than a single ion or a linear chain remains an outstanding challenge in the field and stands to benefit from alternative ways to avoid micromotion. One recently demonstrated approach along these lines relies on the preparation of highly excited Rydberg states within a Bose-Einstein-Condensate, which then behave as ions immersed in an atomic bath \cite{Kleinbach2018}. So far the achieved kinetic energies of the Rydberg atoms correspond to temperatures on the order of microkelvin, and the sensitivity of the system to stray electric fields as well as the comparably short lifetime of the Rydberg excitations remain challenging.\\

In view of these considerations, a naturally seeming approach for addressing the issues highlighted above would be to use the powerful set of tools and methods provided by optical potentials and combine it with the advantaged granted by ions and Coulomb crystals as a platform for experimental investigations. The recently emerged field of optical traps for ions aims at harvesting exactly the potential of this synergetic combination \cite{Schaetz2017}. Dynamically applied optical forces are instrumental in quantum information processing with ions \cite{Wineland2013,Leibfried2003} as a route towards mediating state-dependent interactions \cite{Monroe1995,Blatt2012,Friedenauer2008,Blatt2008}. Static optical potentials have been used to study anomalous ion diffusion \cite{Katori1997}. Despite the broad use of optical forces and potentials for the manipulation of ions, it was only in 2010 that they have been successfully used to demonstrate optical trapping of a single ion \cite{Schneider2010} in absence of any rf fields. Since then the employed techniques and methods have undergone a continuous development in terms of performance and feasibility culminating in a series of experiments \cite{Schneider2012,Schneider2012b,Enderlein2012,Huber2014,Lambrecht2017,Schaetz2017,Schmidt2018}. These investigations will be in the focus of the following discussion aiming  to provide a dissemination of the basic concepts behind this approach as well as its applicability to the outstanding experimental issues outlined in the previous paragraphs.\\

Another closely related class of experiments is based on a combination of radiofrequency traps with optical standing waves \cite{Linnet2012,Karpa2013,Laupretre2019}. This approach provides composite trapping potentials with wavelength-scale periodicity whose implementation in conventional, purely rf-driven, traps including planar chip technology, to this day remains a highly demanding task exceeding the state-of-the-art.
It is worth pointing out that combining several established trapping techniques in order to obtain a system with a unique set of advantageous properties has been a very successful strategy in the past. For example, a combination of linear Paul traps and Penning traps allows for a suppression of the magnetron frequency and can lead to an improved performance of mass spectrometers \cite{Huang1997}.
In the more recent case, the obtained nanoscale corrugation of the effective potential in one dimension is the unique feature of the hybrid setups mentioned above. They turned out to be a nearly ideal platform for investigating theoretically treated phenomena such as structural transitions of one-dimensional ion chains \cite{Pruttivarasin2011,Benassi2011} and have recently lead to the first realization of friction models on the level of a few atoms \cite{Bylinskii2015,Gangloff2015}. Although in this hybrid trapping approach the presence of micromotion still hinders the straightforward extension to higher dimensional systems, in the future, radiofrequency-free trapping potentials realized in optical traps could open up a novel field of studies shedding light on structural phase transitions between one, two and three dimensional ordering close to a quantum critical point \cite{Baltrusch2011,Baltrusch2012,Shimshoni2011,Shimshoni2011a}. This however requires the capability to prepare such higher-dimensional Coulomb crystals in the quantum regime which is another outstanding challenge in contemporary atomic physics.\\

On a similar note, the availability of quasi-static and state-dependent potentials confining 2d ion crystals could also provide a platform for novel quantum simulations \cite{Schneider2012b,Zoller2000}, e.g. the study of fracton models \cite{Pretko2018} or contribute to the field of quantum information processing by utilizing the state-dependent nature of optical potentials to realize fast gate operations as an alternative to approaches based on Rydberg ions \cite{Feldker2015,Higgins2017,Higgins2017a} or fast Raman transitions \cite{Campbell2010}. Lastly, ions in optical potentials may be a used as a nanoscale probe with implications in the field of metrology and sensing of ambient fields. \\

The aim of the following sections is to present an overview of the recent advances in the field of optical trapping of ions and Coulomb crystals as well as to provide a detailed description of the prerequisites, experimental techniques, limitations and prospects of this approach for such envisioned applications.
\chapter{Trapping ions with light fields} \label{chpt:ODTs}

\begin{figure} [h!]
 	\subfigure[]{\includegraphics[width=0.48\textwidth]{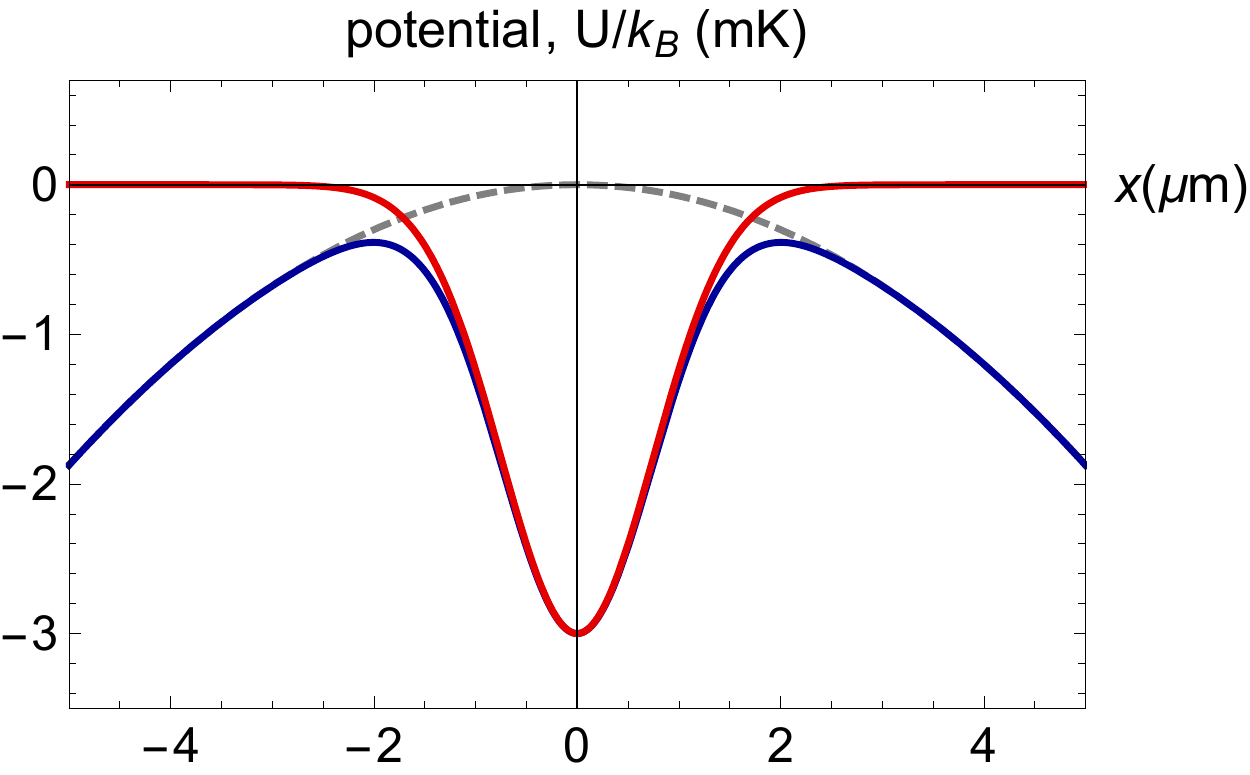}}
  	\hspace{0.5cm}		
 	\subfigure[]{\includegraphics[width=0.48\textwidth]{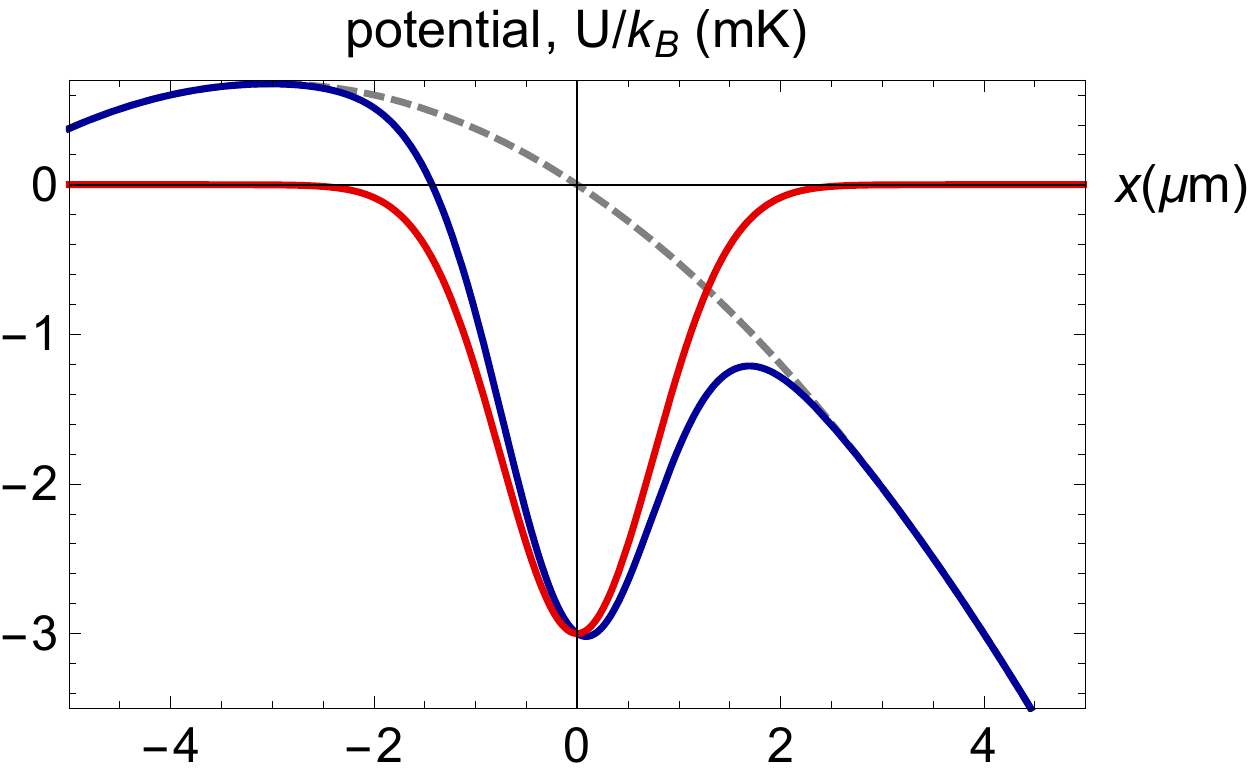}}
 	\caption{\textbf{(a)} 
 	Effective potential $ U/k_B $, with $ k_B $ the Boltzmann constant, (dark blue solid line) experienced by an ion in a repulsive quadratic electric field and an attractive optical potential: the effective trap depth is reduced as compared to the optical trap depth (red solid line) due to the defocusing electrostatic potential (grey dashed line) along the $ x $-axis. \textbf{(b)} Accounting for a finite stray electric field leads to an additional tilt of the effective potential, further reducing its depth.}
 	\label{fig:PotStatOptDeconf}	
 \end{figure}

In principle, the optical forces experienced by ions are of the same origin as for the case of neutral atoms. This is due to the fact that the light shift responsible for the optical potential stems from the coupling of the light field to the outer electron of the atoms and not to the electric charge. Thus, in a field-free environment, the effective potentials generated by an optical field for an ion and a neutral atom of the same polarizability and energy level structure would be identical. So in order to obtain a trapping potential for an ion, we would only have to replace the potential generated by an external electric quadrupole field modulated at radiofrequency with suitable optical fields, e.g. a red-detuned Gaussian beam focused on the ion. In reality however, external electric fields are always finite, and in order to satisfy Laplace's equation,
\begin{equation}
\nabla^2 \Phi = 0
\end{equation}
 where the quadratic potential of a quadrupole field is determined by $ \Phi (x,y,z)$ $ \propto $ $\alpha x^2 + \beta y^2 + \gamma z^2 $ with the factors $ \alpha, \beta, \gamma $ obeying the condition $ \alpha + \beta + \gamma = 0 $ \cite{Leibfried2003}, they cannot be arranged in a way to provide static confinement in all three dimensions.
 This is equivalent to a formulation found in Earnshaw's theorem: \enquote{An electrified body placed in a field of electric force cannot be in stable equilibrium} \cite{Maxwell1873}. Laplace's equation imposes restrictions on the possible solutions $ \Phi $ to the effect that no local extrema in free space are allowed, whereas local saddle points are possible.
 In other words, the curvature of a purely electrostatic potential is negative in at least one direction. To illustrate this aspect, let us consider an ion placed at the point of unstable equilibrium, i.e. maximum of a quadratic electric field with $ \beta, \gamma >0  $ and $ \alpha < 0 $. This point coincides with the node of the radiofrequency field in a perfectly compensated Paul trap. For simplicity, let us also assume that the direction of the resulting deconfinement coincides with the $ x $-axis as shown in figure \ref{fig:PotStatOptDeconf}. In presence of an optical trapping field originating from a Gaussian, red-detuned beam propagating along the $ z $-axis, focused on the ion in a hypothetical scenario without any stray electric fields $( E_{S} = 0 )$, an ion at $ x = 0 $ will be subject to a trapping optical potential with a reduced depth due to the negative curvature of the electric field. This fundamental reduction of the effective trap depth is drastically aggravated if an additional linear stray electric field is added, as illustrated in figure \ref{fig:PotStatOptDeconf}(b).

While in realistic experiments the quadratic component of the electric field is determined by the configuration of electrostatic electrode parameters of the Paul trap, the linear contribution originates from uncompensated stray electric fields and is responsible for excess micromotion observed in all Paul traps. Already in the case of conventional ion traps great effort is invested in order to compensate such external fields and to minimize their detrimental impact on the performance of ion trap experiments.\\

\begin{figure}
		\begin{center}
		\includegraphics[width=0.6\textwidth]{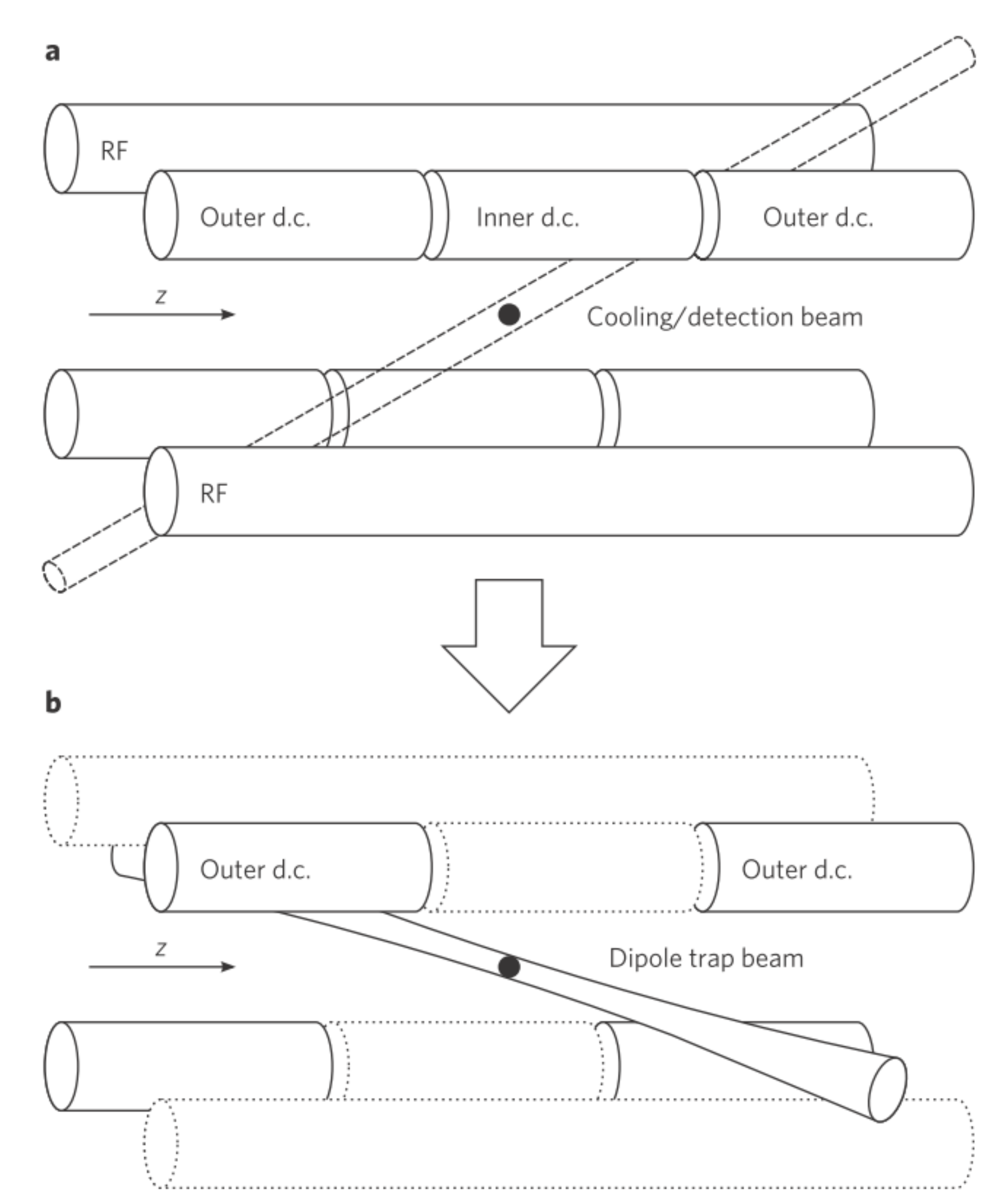}\\
		\end{center}
	\caption{Experimental setup used for the first realization of an optical trap for a single $ {}^{24}\mathrm{Mg}^+ $ ion. (a) Initialization of a laser-cooled ion in a Paul trap with radial and axial confinement provided by rf electrodes and voltages applied to the d.c. electrodes, respectively. (b) During the optical trapping phase, the rf potentials are turned off. Radial confinement is now achieved with an optical dipole trap. Taken from \cite{Schneider2010}.}
	\label{fig:OptTrapFirst}
\end{figure}

In the case where we seek to replace the confinement provided by radiofrequency fields with optical forces, it is instructive to compare the available optical forces with those generated by uncontrollably deposited charges on the electrodes of a typical Paul trap. Assuming a corresponding potential of $ 1 V $ and ion to electrode distances of $ 1 \,  \mathrm{mm} $, we obtain electric forces on the order of $F_{el,str} \sim 10^{-17} \, \mathrm{N}$, whereas with commercially available laser sources optical forces of about   $F_{opt} \sim 10^{-19} \, \mathrm{N}$ can be achieved. Under such conditions Coulomb forces are orders of magnitude larger than optical forces, and trapping ions with optical fields would not be feasible. The conclusion of these considerations is that the obtainable performance of optical ion traps will hinge on the capability to detect and compensate stray electric fields as precisely as possible. Whether or not this can be done at a level sufficient for optical trapping was unclear until the first proof-of-principal demonstration of optical trapping of a single ion in 2010 \cite{Schneider2010}. The compensation accuracy achieved in this experiment, schematically shown in figure \ref{fig:OptTrapFirst}, was on the order of $ | \delta E_{S} | \sim 100 \, \mathrm{mV /m} $, corresponding to residual electric forces of $F_{el,str} \sim 10^{-20} \, \mathrm{N}$.

In the following, we will discuss the prerequisites and experimental techniques for carrying out such experiments, followed by the description of a generic protocol for optical trapping.

\section{Prerequisites for optical trapping experiments}

One of the basic prerequisites in all optical trapping experiments involving single atoms is a sufficient isolation from the environment. In contrast to optical tweezers experiments for small macroscopic objects, any collision with background gas leads to atom loss from the trap due to the comparatively small mass and typically low trap depths in such experiments. Most commonly, optical traps for atoms are realized under ultra-high vacuum conditions, with pressures of the background gas well below $ 10^{-9} $ mbar, resulting in collision rates below $ 1 \, \mathrm{s}^{-1} $ \cite{Grimm2000}. Analogous to the case of neutral atoms, in general the necessary preparation, manipulation and detection relies on the availability of electronic transitions allowing for laser cooling and imaging. The latter is crucial for the stray electric field compensation schemes described in this work. This condition is met for a number of routinely employed ion species like $ \mathrm{Be}^+ $, $ \mathrm{Mg}^+ $, $ \mathrm{Ca}^+ $, $ \mathrm{Zn}^+ $, $ \mathrm{Sr}^+ $, $ \mathrm{Cd}^+ $, $ \mathrm{Ba}^+ $, $ \mathrm{Yb}^+ $, $ \mathrm{Lu}^+ $, $ \mathrm{Hg}^+ $ which have been laser cooled in Paul traps. While the capability to sympathetically cool ions that are not resonant with any of the present optical fields with high efficiency by embedding them in a laser cooled Coulomb crystal is an advantage of ions over neutral atom systems, this kind of experiments requires a substantially more refined level of control that has only been demonstrated recently \cite{Schmidt2018}.\\

As we will see in the following, the presence of a metastable electronic level can be used to obtain trapping potentials for ions that directly depend on the electronic state. This is the case not only for several commonly used elements, e.g. $ \mathrm{Yb}^+ $, $ \mathrm{Ba}^+ $, $ \mathrm{Ca}^+ $, $ \mathrm{Sr}^+ $ but also for more exotic species such as $ \mathrm{Lu}^+ $ or $ \mathrm{Eu}^+ $.
 The focus of this chapter will be on the simplest possible realizations of optical trapping of ions. A detailed discussion of the more advanced experiments with ion crystals will be presented later on. The fact that in the described experiments optical forces acting on the ions have to overcome their Coulomb interactions with ambient electric fields gives rise to a specific set of prerequisites.\\

As outlined in the previous section, the configuration of the electrostatic fields has a crucial impact on the effective trapping potential. In particular, any confining potential results in a negative curvature for at least one of the orthogonal directions, reducing the effective trap depth. Since it is advantageous to maintain a controlled static confinement, e.g. along the linear axis of commonly used Paul traps, a convenient starting point is to adjust the axial curvature as low as possible in order to stay within the stability region of the trap and to ensure efficient laser cooling. This can lead to secular frequencies of approximately $ \omega_{ax} = 2 \pi \times \, 10 \, \mathrm{kHz} $, which is at least an order of magnitude lower than the secular frequencies encountered in typical ion trapping experiments relying on a large separation of the vibrational levels, at least during the preparation phase. For such axial confinement settings, the negative curvature due to the electrostatic potential easily becomes comparable to the curvatures provided by even strongly focused high power lasers. In a second step, the radial compensation electrodes can be adjusted such that the resulting deconfinement is distributed equally among the two orthogonal radial directions. For a focused Gaussian beam propagating along the nodal line of the Paul trap this corresponds to a symmetric reduction of the effective radial potential and maximizes the trap depth for a given optical power in the dipole trap.\\

While in principle the detrimental consequences of the Coulomb interaction in an optical ion trapping experiment can be overcome by using sufficiently powerful laser sources, the vast majority of experiments largely benefits from a precise compensation of stray electric fields. It was shown that an accuracy level of approximately $ | \delta E_{S} | \approx 100 \, \mathrm{mV / m} $ is sufficient for proof-principle experiments carried out with commercially available laser sources \cite{Schneider2010,Enderlein2012} and for certain applications such as controlled preparation of a predefined number of ions which will be discussed in the outlook section. On the other hand, applications in the field of ion-atom interactions require minute control over the effective potentials enabled by compensation uncertainties on the order of $ | \delta E_{S} | =  10 \, \mathrm{mV / m} $ and below \cite{Cetina2012}. \\

As we will see in the following, the most straightforward way to fulfilling these requirements is to use a linear Paul trap in the outlined configuration for the preparation and detection steps. Paul traps with a large ion-electrode distance of several $\mathrm{ mm} $  are particularly suited due to their compatibility with high numerical aperture optics crucial for precise stray field sensing and compensation, high resolution of the required electrostatic potentials, and intrinsically low anomalous heating rates. Nonetheless, most of the mentioned techniques are also applicable for planar traps, e.g. interfaced with build up cavities as demonstrated in recent works utilizing a combination of radiofrequency and periodic optical fields \cite{Karpa2013,Bylinskii2015,Gangloff2015}.

\section{Methodology} \label{sec:Methods}

This section provides an overview over the most useful methods commonly employed for optical trapping of ions. These cover different aspects ranging from loading and suitable configurations of ion traps and preparation of the ions for an optical trapping sequence, to detection and analysis.

\subsection*{Paul trap}

A linear Paul trap with comparatively large ion to electrode distances of several millimeters such as the one depicted in figure \ref{fig:PaulTrapAndIons}(b) provides the following advantages in view of storing ions in dipole traps:\\

\begin{itemize}

\item Firstly, it provides a large trapping volume, allowing for efficient ion loading. Schemes involving photo-ionization with transitions in the ultraviolet spectral range, e.g. light at $ 405 \, \mathrm{nm} $ used for the ionization of electronically excited $ \mathrm{Ba} $ atoms, benefit from a reduced exposure time, mitigating the deposition of charges on dielectric surfaces such as windows or cavity mirrors. Continuous operation of an oven has been shown to produce increased drift rates for stray electric fields, such that ablative loading schemes bear the potential for improved performance of optical trapping experiments.

\item Large numerical aperture optics allow for focussing of dipole trapping beams to small beam waists. For a given optical trap depth, the restoring force increases with smaller beam radii. Thus, smaller dipole traps provide more robustness against forces from stray charges. The same optical access can be used to collect ion fluorescence which, in combination with high spatial resolution of the ion's position, facilitates highly accurate stray field compensation \cite{Berkeland1998,Huber2014}.

\item High resolution of electrical potentials applied for compensation. With commercially available digital-to-analog converters, field resolutions on the order of $ 1 \, \mathrm{mV / m} $ are readily achievable without the necessity for additional passive components such as voltage dividers which can alter the response of the system in a dynamical regime of operation.

\item The so-called anomalous heating is the subject of active investigations, and has been shown to exhibit a strong dependence on the ion-to-electrode distance $ d $, with a typical scaling of approximately $ d^{-4} $ \cite{Deslauriers2006,Turchette2000}. While in the case of planar traps with $ d \sim 30 \, \mu \mathrm{m}$ extensive measures have to be taken in order to ensure heating rates on the order of one vibrational quantum per $ \mathrm{ms} $ \cite{Hite2012}, electrode-to-ion distances in the $ \mathrm{mm} $ range provide several orders of magnitude lower heating rates. This is beneficial for example in the case of future investigations of ion-atom-interactions in the quantum regime, and for potential applications in the field of quantum simulations with ions.
\end{itemize}
At the same time, large Paul traps invoke several disadvantages that can affect the performance of optical trapping experiments. \\

\begin{itemize}
\item A prominent example is the relatively low secular frequency of the radiofrequency confinement, making established methods for ground state cooling more challenging to implement.
\item Maximized optical access can come at the expense of reduced shielding from ambient stray electric fields stemming from charges deposited on windows or other dielectric surfaces.
\item The downside of a large trapping volume is its susceptibility to highly energetic ions stored on large trajectories within the trap that lead to heating and reduce the efficiency of experiments requiring a predefined number of ions due to random crystallization. The latter effect can be effectively mitigated once optical trapping conditions have been established, since the trapping region is then restricted to a much smaller volume given roughly by $ V_{opt} \approx 2 \pi w_0^2  z_R$, with $ w_0 $ being the waist radius and $ z_R $ the Rayleigh range of the dipole trapping beam. 

\end{itemize}

\subsection*{Electrostatic field configuration}

The effective potential for a single ion in a focused beam dipole trap in a direction $ i \in \left\{ x, y, z \right\} $ is a superposition of an electrostatic potential $$ U_{dc,i} = \frac{1}{2} m \omega_i^2 i^2 + e E_{str, i} i $$ comprising the attractive or repulsive contributions stemming from potentials applied to the Paul trap electrodes combined with the stray electric field $ \vec{E}_{str} $, and the purely optical potential $$ U_{opt}(r,z) = U_0 \frac{\exp(-2 r^2 / w(z)^2)}{1 + (z / z_R)^2} $$ as illustrated in figure \ref{fig:PotStatOptDeconf}:
\begin{equation}
U_{tot} = U_{dc} + U_{opt}.
\label{eqn:TotalPotential}
\end{equation}
Here, $ r, z, z_R, w(z) $ denote the radial and axial coordinates, the Rayleigh range and beam radius, respectively. Therefore, the configuration of electrostatic fields is an important aspect that has to be taken into account in order to achieve optimal performance of optical traps. This configuration strongly depends on the parameters of the experiment and in particular on the chosen dipole trap geometry.\\

In order to mitigate the reduction of the effective potential depth along the directions of defocusing due to the applied electrostatic fields, it is helpful to use the following strategy: firstly, during the optical trapping phase the focusing potential contribution $ U_{dc} $ should be kept as weak as possible, and secondly, the resulting defocusing should be distributed equally along the two remaining orthogonal directions. This requires a method for measuring secular frequencies as well as a set of electrodes sufficient for tuning of the respective electrostatic potentials. Secular frequencies can be determined with an accuracy of at least $ 0.1 \, \mathrm{kHz} $ by resonantly exciting laser cooled ions with modulated external electric potentials while monitoring their fluorescence and spatial extent e.g. with a charge-coupled device (CCD) camera. If the excitation occurs at the secular frequency of the ion, a distinct expansion can be observed. In recent experiments with focused beam dipole traps, typical values between $ \omega_{dc} = 2 \pi \times 10 \, \mathrm{kHz} $ and $ 2 \pi \times 20 \, \mathrm{kHz} $ have proven to be a suitable choice. \\
 
In future experiments aiming to use the technique of optical trapping for confining 2d ion Coulomb crystals a different trapping geometry is advantageous. A straightforward looking way to approach this problem would be to increase the axial confinement of an initially linear chain to a point where a structural transition to a zig-zag configuration and later on to a 2d crystal takes place while simultaneously increasing the optical intensity of the dipole trap to compensate the resulting defocusing from the electrostatic field and mutual Coulomb repulsion between the ions. This approach invokes the disadvantage of having to use comparatively large beam waists to fit in the Coulomb crystals and correspondingly very large optical powers that easily can be in excess of commercially available solutions. Instead, one might use a standing wave geometry to provide the axial confinement, and use electrostatic potentials for radial confinement. This arrangement should provide both a sufficient robustness with respect to axial defocusing as well as stray electric fields due to a much stronger gradient of the optical potential achievable in optical potentials with a $ \lambda $-scale periodicity. This aspect is also discussed in section \ref{chpt:highDimCrystals}.

\subsection*{Stray electric field compensation} \label{sec:StrayComp}

\begin{figure} [h!]
	\begin{center}
	\includegraphics[width=0.7\textwidth]{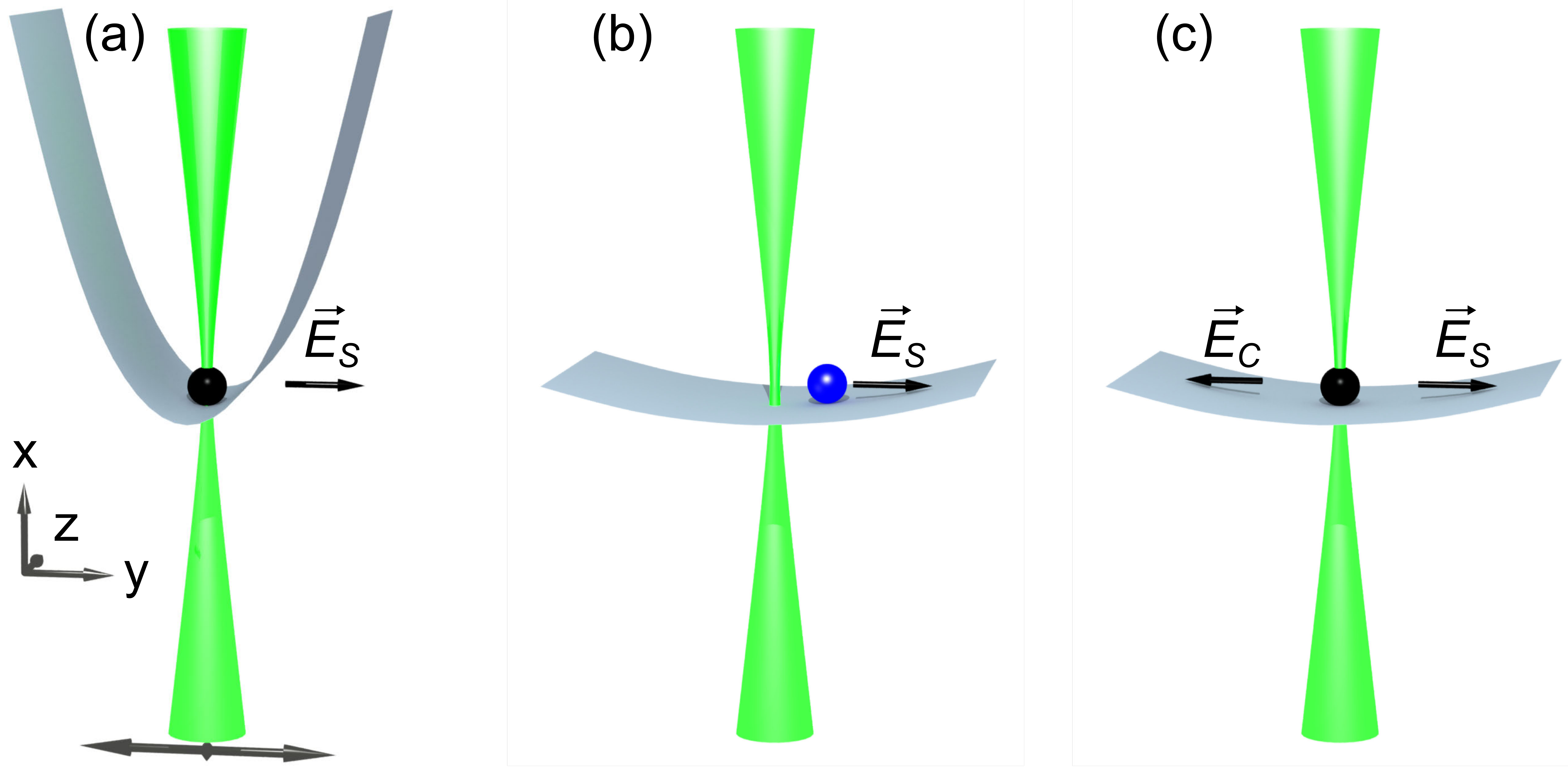}
	\end{center}
	\caption{ Principle of the a.c. Stark shift-assisted stray electric field compensation scheme. A residual stray electric field $ E_S $ causes a displacement of the ion in (a) a harmonic rf pseudopotential with stiff confinement (gray parabolic surface) and shallow confinement (b,c). The initial displacement is amplified by the factor $ \left( \omega_{rf,h} / \omega_{rf,l} \right)^2 $. Compensation potentials are then used to minimize the displacement resulting in a residual uncertainty of $| \delta E_{S} | < 9 ~\mathrm{mV/m} $. Taken from \cite{Huber2014}. }
	\label{compscheme}
\end{figure}

Since the rf fields are switched off during the optical trapping period, the ion does not experience micromotion. On the other hand, the same methods applied for minimizing excess micromotion are also effective in reducing the influence of stray electric fields. While several methods have been developed over the last decades, in the described setting an approach based on measuring the displacement of an ion exposed to a static stray field at different rf confinements with secular frequencies $ \omega_{rf,h} $ and $ \omega_{rf,l} $, as illustrated in figure \ref{compscheme}, has proven particularly sensitive. This method is an adapted version of ideas described in 1998 by D. J. Berkeland and colleagues \cite{Berkeland1998}: \enquote{In the first of these methods, which is sensitive to excess micromotion caused by static fields, the time-averaged ion position is monitored as the pseudopotential is raised and lowered}.

In this scheme, the achievable stray field detection accuracy scales as the square of the ratio of the secular frequencies, i.e. $ \delta E_{S} \propto \left( \frac{\omega_{rf,l}}{\omega_{rf,h}} \right)^2 $, such that the capability to switch between a stiff and a shallow rf-potential, e.g. with $ \omega_{rf,h} \approx 2 \pi \times 300 \, \mathrm{kHz} $ and $ \omega_{rf,l} \approx 2 \pi \times 30 \, \mathrm{kHz} $, is highly advantageous. Using the light shift induced by the focused dipole trap beam as a marker for the node of the rf field and as a way to derive position when moving an ion through the beam, a residual uncertainty of $ |\delta E_{S}| = 9 \, \mathrm{mV/m} $ has been demonstrated, with the corresponding measurements being carried out within about $ 90 \, \mathrm{s} $ \cite{Huber2014}. 
Apparatuses with sufficiently high spatial resolution of the imaging system in the radial directions benefit from the capability to directly monitor the ion position for different rf confinement settings. This scheme yields an accuracy between $ 5 \, \mathrm{mV/m} $ and $ 10 \, \mathrm{mV/m} $ after a measurement time of about $ 30 \, \mathrm{s} $ with a position resolution of $ 200 \, \mathrm{nm} $.
\subsection*{Temperature measurement}
\begin{figure}[h!!]
\centering
		\subfigure[]{\includegraphics[height=0.29\textwidth]{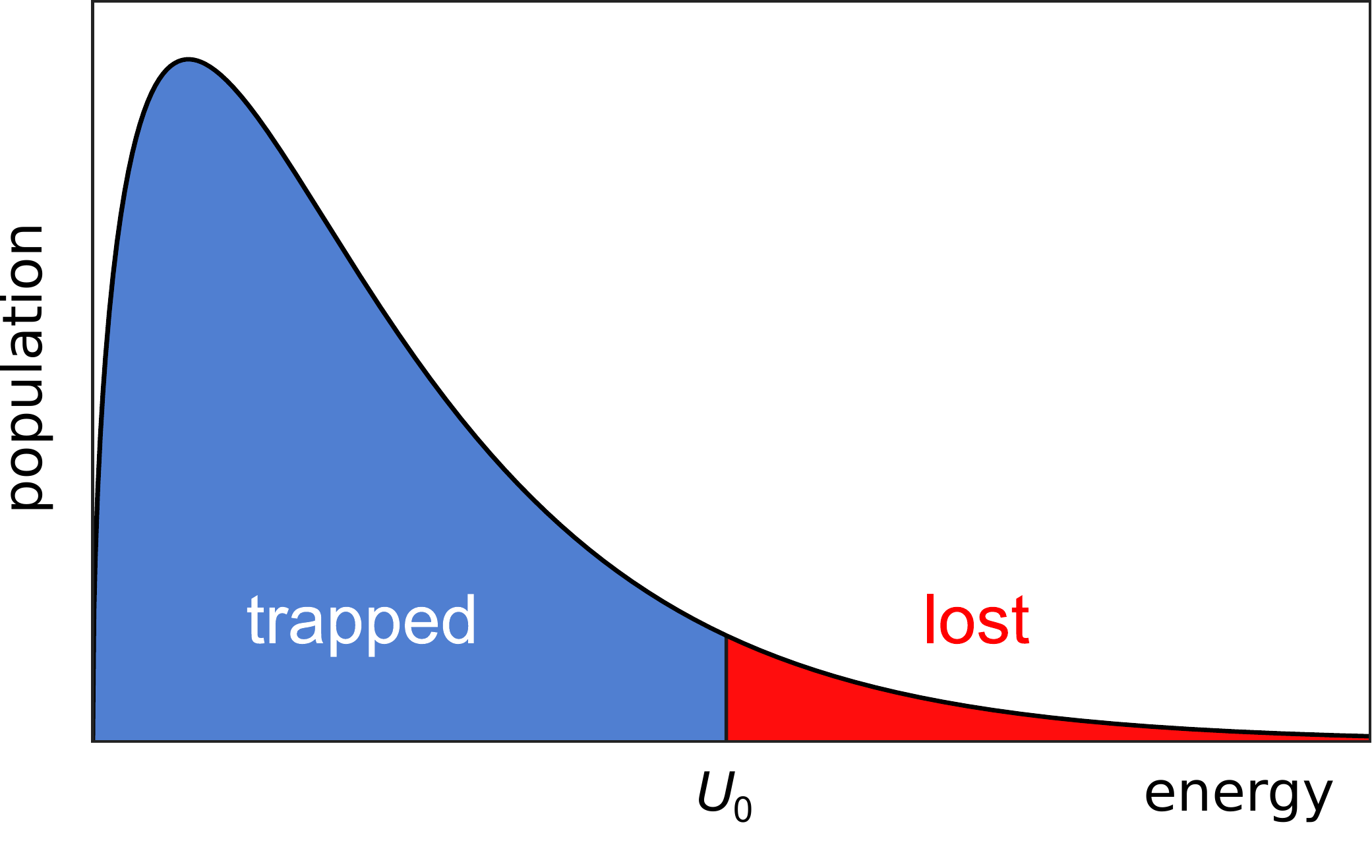}
			\label{fig:BoltzmannTruncated}}
		\subfigure[]{
		\includegraphics[height=0.3\textwidth]{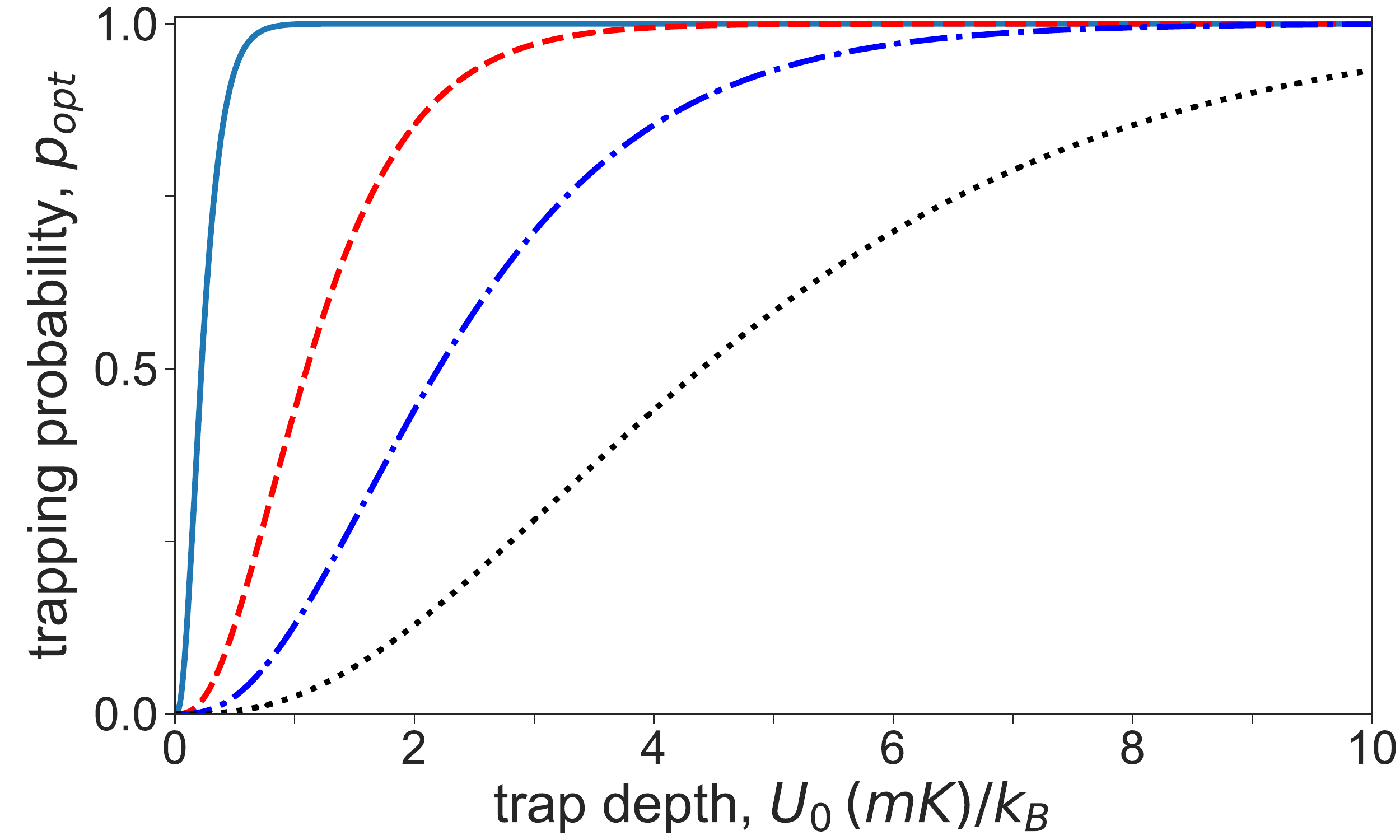}
	\label{fig:BoltzmannT}}
	\caption{Illustration of expected trapping probability $ p_{opt} $ in a trap of depth $ U_0 $ according to a simplified model considering a truncated Maxwell-Boltzmann distribution. \textbf{(a)} The ratio of the accumulated populations in all energy levels below $ U_0 $ and above the threshold determines the expected probability for a successful experimental realization. \textbf{(b)} Examples of expected trapping probabilities based on the radial cutoff model for ions prepared at initial temperatures of $ 0.1 \, \mathrm{mK} $ (solid line), $ 0.5 \, \mathrm{mK} $ (dashed line), $ 1.0 \, \mathrm{mK} $ (dash-dotted line), and $2.0 \, \mathrm{mK} $ (dotted line) as a function of applied trap depth $ U_0 $. }
	\label{fig:Boltzmann}
\end{figure}
The trapping probability of an atom in a potential with depth $ U_0 $ depends on the initial temperature of the atom $ T $ as well as on the trap depth. In the simplified model based on a truncated Boltzmann distribution as depicted in figure \ref{fig:BoltzmannTruncated} the probability $ p_{en} $ for a successful experimental realization is given by the following expression \cite{Tuchendler2008}:
$$ p_{en}(\xi) = 1- \left( \frac{\xi^2}{2} + \xi + 1 \right) e^{-\xi} ,$$
with $ \xi = \frac{U_0}{k_B T} $. This simplified picture neglecting energy stored in the angular momentum of the atoms can be extended to a more complete radial cutoff model which yields a modified expression for the survival probability \cite{Schneider2012}:

\begin{equation}
p_{rad}(\xi) = 1- e^{-2 \xi} - 2 \xi e^{-\xi}.
\label{eqn:radialcutoff}
\end{equation}

According to this result, the temperature of an atom can be determined by measuring the survival probability for different trap depths as illustrated in figure \ref{fig:BoltzmannT}. This method provides the advantage of measuring the temperature for a wide range of trapping durations $ \tau $ directly within the trap without additional manipulation of the atoms. However, since the actual energy of the atom can be fixed during the trapping phase and the outcome of a trapping attempt is binary, this approach requires accumulation of statistics by repeating the experiment several times in order to obtain a sufficiently accurate representation of the energy distribution.

\subsection*{Optical trapping protocol}

\begin{figure}[h!]
		\begin{center}
		\includegraphics[height=0.3\textwidth]{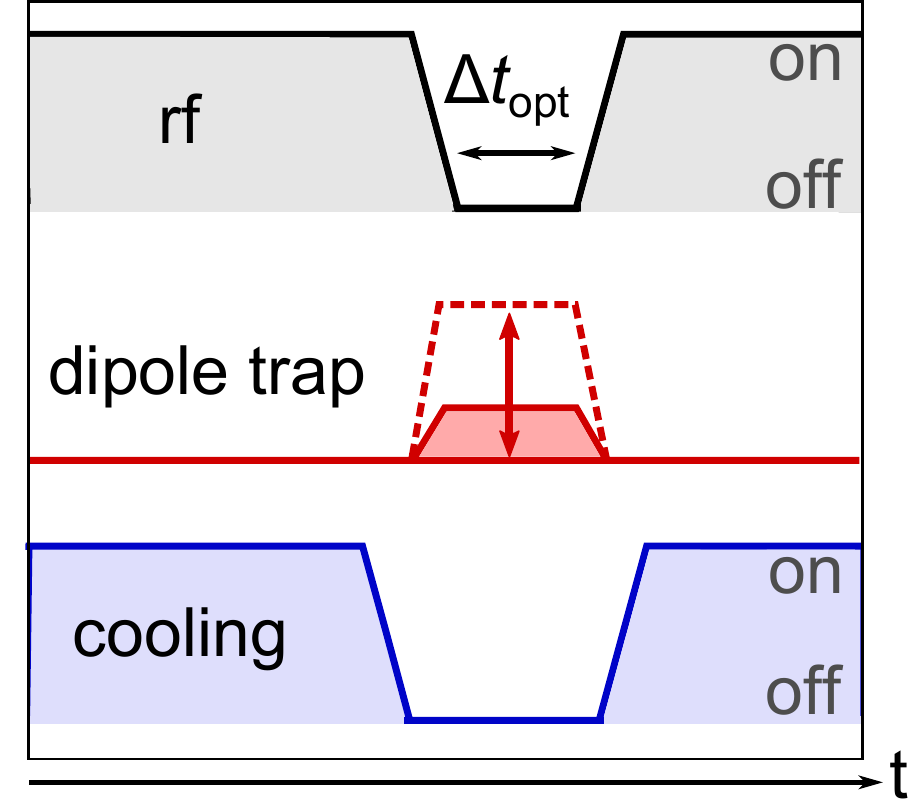}	
		\end{center}
	\caption{
	The experimental protocol comprises three phases: (1) Preparation: loading and Doppler cooling the ions, stray field compensation; (2) Trapping: transfer into the dipole trap by turning off the rf field and cooling lasers for the optical trapping duration; (3) Detection: ion fluorescence is used to determine if the trapping attempt was successful.
	}
	\label{fig:OptTrapSequence}
\end{figure}

A generic sequence suitable for optical trapping of ions, schematically represented in figure \ref{fig:OptTrapSequence}, comprises the following steps subdivided in three phases:\\

\begin{itemize}
\item Phase 1: preparation

	\begin{itemize}
		\item{loading into Paul trap}
		\item{Doppler cooling}
		\item{stray field compensation}
	\end{itemize}		
		
\item 	Phase 2: transfer and trapping
	\begin{itemize}
		\item{ramp optical dipole trap to $ U_0 $}
		\item{ramp rf field amplitude to zero (not dc potentials)}
		\item{trap ion optically for a duration $\tau$}
	\end{itemize}	
		
\item	Phase 3: detection
	\begin{itemize}
		\item{switch Paul trap on}
		\item{switch dipole trap off}
		\item{fluorescence detection}   
	\end{itemize}

\end{itemize}
	
The phases 1 and 3 of the optical trapping sequence are straightforward in view of the previously described methods. The transfer between the rf and optical traps however requires taking certain precautions, since the Paul trap is operated outside of the optimized settings within the stability diagram and ultimately is rendered unstable, at which point the optical confinement has to take over. A critical parameter is the ramp duration of the rf potential and the dipole trap, respectively. It has to be chosen carefully in order to maintain adiabaticity while applying the changes fast enough to mitigate the detrimental effects of crossing instabilities encountered in rf traps with higher order anharmonicities \cite{Alheit1995}. At radial secular frequencies on the order of $ \omega_{rf,rad} \approx 2 \pi \times 100 \, \mathrm{kHz} $ typical ramp durations of about $ 50 \, \mu \mathrm{s} $ have been shown to provide low additional heating and reliable results.

\subsection*{Optical trapping probabilities}
Optical ion trapping attempts are realizations of Bernoulli experiments and as such yield inherently binary results. With the detection being based on observing the fluorescence of an ion after a trapping attempt, the achievable fidelities can approach $ 100 \% $. The ratio of successful and total attempts $ N_S / N_T $ can be used to calculate the most likely underlying probability. The uncertainties of such measurements can then be estimated by calculating corresponding Wilson score intervals \cite{Wilson1927}, and are purely statistical. The exact value of the uncertainties depends on $ N_S / N_T $, but typically, in order to obtain uncertainties of about $ \pm 15 \% $, approximately $ 20 $ attempts are required.
\chapter{Optical dipole traps for single ions} \label{chpt:SingleIonODTs}

\begin{figure} [h!]
	\begin{center}
	\includegraphics[width=0.5\textwidth]{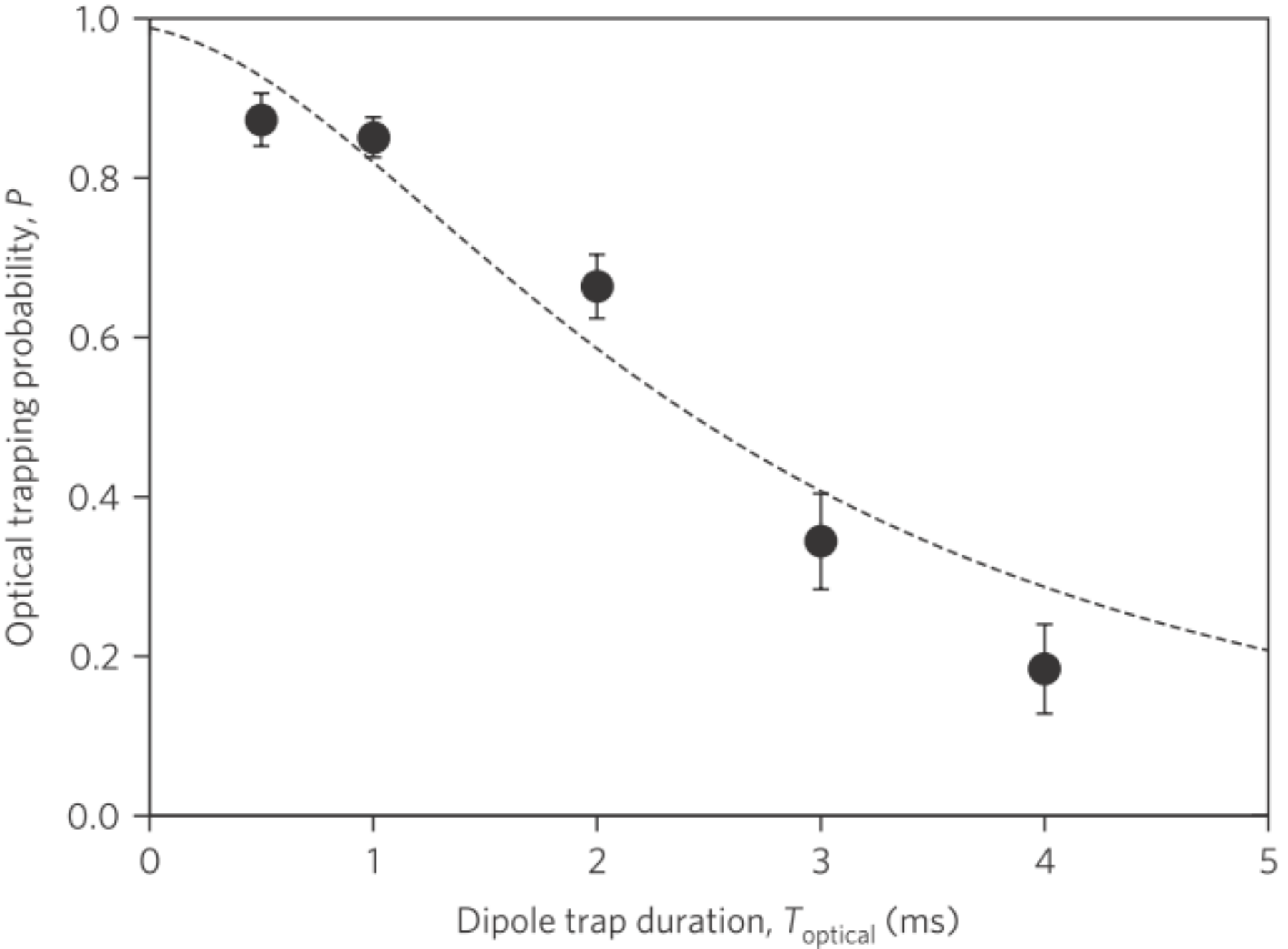}
	\end{center}
	\caption{ Optical trapping probability of a $ ^{24} \mathrm{Mg}^+ $ ion in a near-resonant optical dipole trap as a function of the trapping duration. Taken from \cite{Schneider2010}.}
	\label{Mg_opttrap_lifetime}
\end{figure}

The first demonstration of optical trapping of an ion was achieved in 2010, where a near resonant focused dipole trap was used to confine a $ \mathrm{Mg}^+ $ ion for a few milliseconds \cite{Schneider2010}. As discussed in the previous chapters, already at this early stage fairly precise compensation of stray electric fields was essential. In order to obtain trap depths of several $ k_B \times 10  \, \mathrm{mK} $, sufficient for optical trapping of ions laser cooled to about $ 5 \, T_D $, a focused beam with a detuning of about $ \Delta \approx \, 6.6 \times 10^3 \, \Gamma $ was used, where $ \Gamma $ is the natural line width of the addressed optical transition and the Doppler cooling limit is denoted with $ T_D \approx 1 \, \mathrm{mK} $. It was found that the heating caused by off-resonant scattering from the dipole trap experienced by the ion within trapping durations of a few $ \mathrm{ms} $ was on the order of the trap depth and thus the main mechanism limiting the observed ion lifetime as shown in figure \ref{Mg_opttrap_lifetime}.\\

Later on, this experiment was successfully carried out within a standing wave of the same wavelength \cite{Enderlein2012}. The lifetimes in this experiment were limited to approximately $ 100 \, \mu \mathrm{s} $, due to the increased off-resonant scattering in the standing wave and due to limitations during the transfer between the rf and optical traps.\\

A similar approach was used in a different set of experiments combining radial rf confinement with an optical standing wave along the third direction. There, the influence of the optical potentials was also demonstrated by observing pinning \cite{Linnet2012} or strongly suppressed ion transport \cite{Karpa2013}. In the latter work it was also shown that the same techniques can be extended to confine small chains with up to 3 ions for at least $ 10 \, \mathrm{ms} $, while simultaneously cooling them close to the ground state and monitoring their individual positions with sub-wavelength resolution. The lifetime of the ions in the 1d standing wave potential was mainly limited by externally modulated electric fields causing additional heating. An extrapolation to the unperturbed case where the ions are left in the dark state shows that the expected lifetime can reach several seconds, unless other fundamental limitations inhibit trapping on these timescales \cite{Turchette2000,Cormick2011}.\\

A very useful property of dipole traps for the simplified case of a two-level atom is the scaling of the potential depth as:
\begin{equation}
U_0(r) =  I(r) \left(\frac{\Gamma}{\Delta} \right) \frac{3 \pi c^2}{2 \omega_0^3},
\label{eqn:Uopt}
\end{equation}
where $ c $ is the speed of light, $ \omega_0 $ the resonance frequency of the electronic transition, and $ I $ is the intensity of the beam at the location of the atoms, while the off-resonant scattering rate $ \Gamma_{sc} $ is given by:
\begin{equation}
\Gamma_{sc}(r) =  I(r) \left(\frac{\Gamma}{\Delta} \right)^2 \frac{3 \pi c^2}{2 \hbar \omega_0^3}
\label{eqn:GammaOpt}
\end{equation}
\cite{Grimm2000}. Thus, for a constant trap depth, the influence of off-resonant scattering is suppressed by trapping atoms in far-detuned dipole traps operated at high intensities.\\

A derivation of the equations \ref{eqn:Uopt} and \ref{eqn:GammaOpt} can be found in contemporary textbooks on atomic physics, e.g. \cite{Foot2005}, and in \cite{Grimm2000}. It should be noted that these relations hold for the case of a two-level atom interacting with far-detuned laser radiation such that the rotating wave approximation is valid. In practice, the trapping light fields interact with multilevel atoms such that the contributions from different electronic states have to be taken into account. For instance, the presence of degenerate magnetic levels of the $ D $-manifolds of $ \mathrm{Ba}^+ $ can lead to a strongly reduced scattering rate on the repumping transition \cite{Berkeland2002} due to population of dark states. By extension, the trapping potential of ions shelved into one of the $ D $-manifolds is expected to exhibit a strong dependence on the composition of the final state. However, the conclusion that the off-resonant scattering rate can be reduced by increasing the detuning still holds in the generalized case, since the effective light shift is essentially a sum of the individual energy shifts from all of the coupled electronic states after accounting for the respective detunings and line strengths. Accordingly, the equations \ref{eqn:Uopt} and \ref{eqn:GammaOpt} are applicable to each of these individual transitions.

\section{Far-off-resonance optical traps for ions}

\begin{figure} [h!]
	\subfigure[]{
	\includegraphics[height=0.4\textwidth]{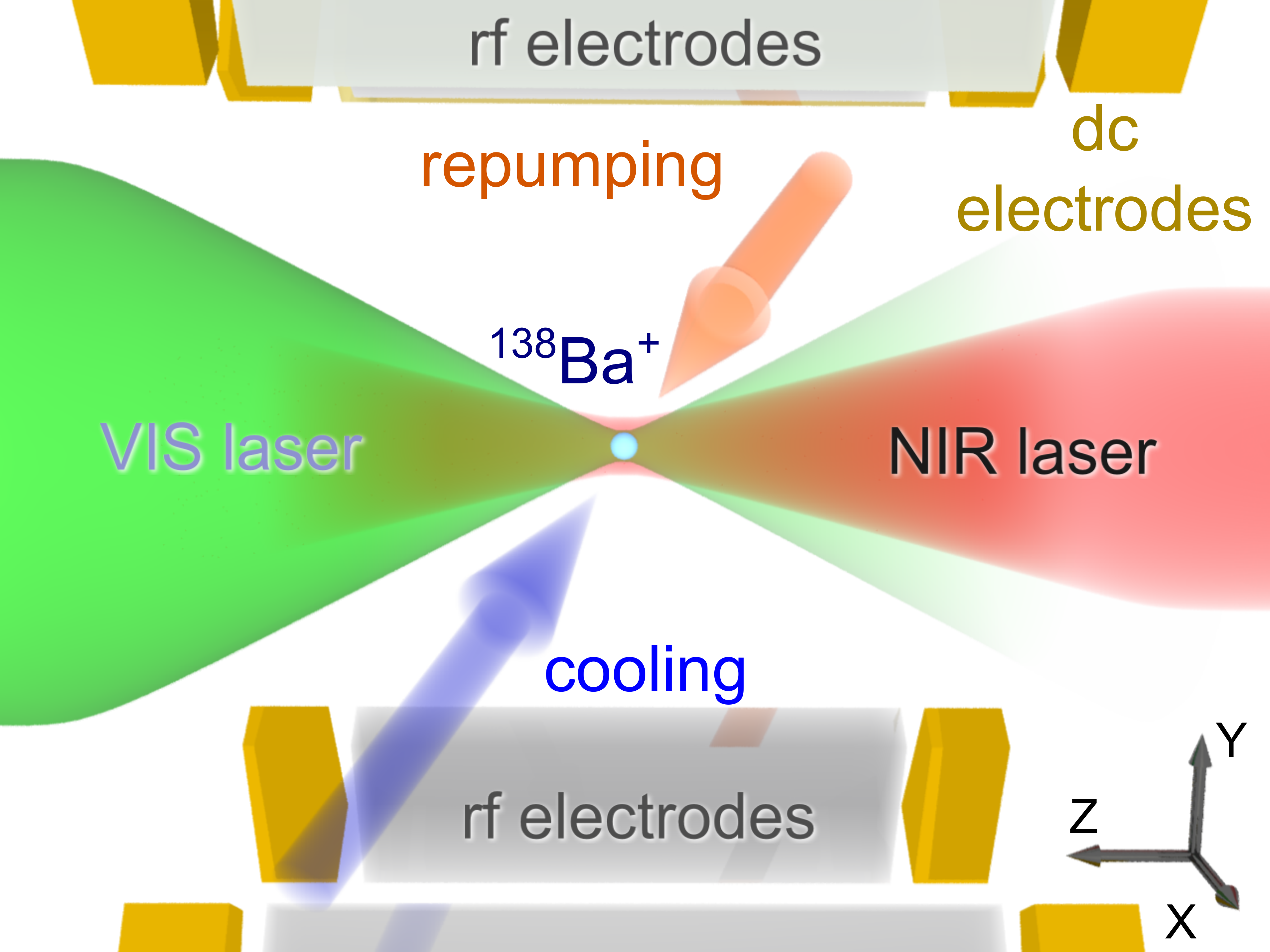}
		\label{fig:LifetimeSetup}}
	\hspace{0.5cm}
	\subfigure[]{
	\includegraphics[height=0.35\textwidth]{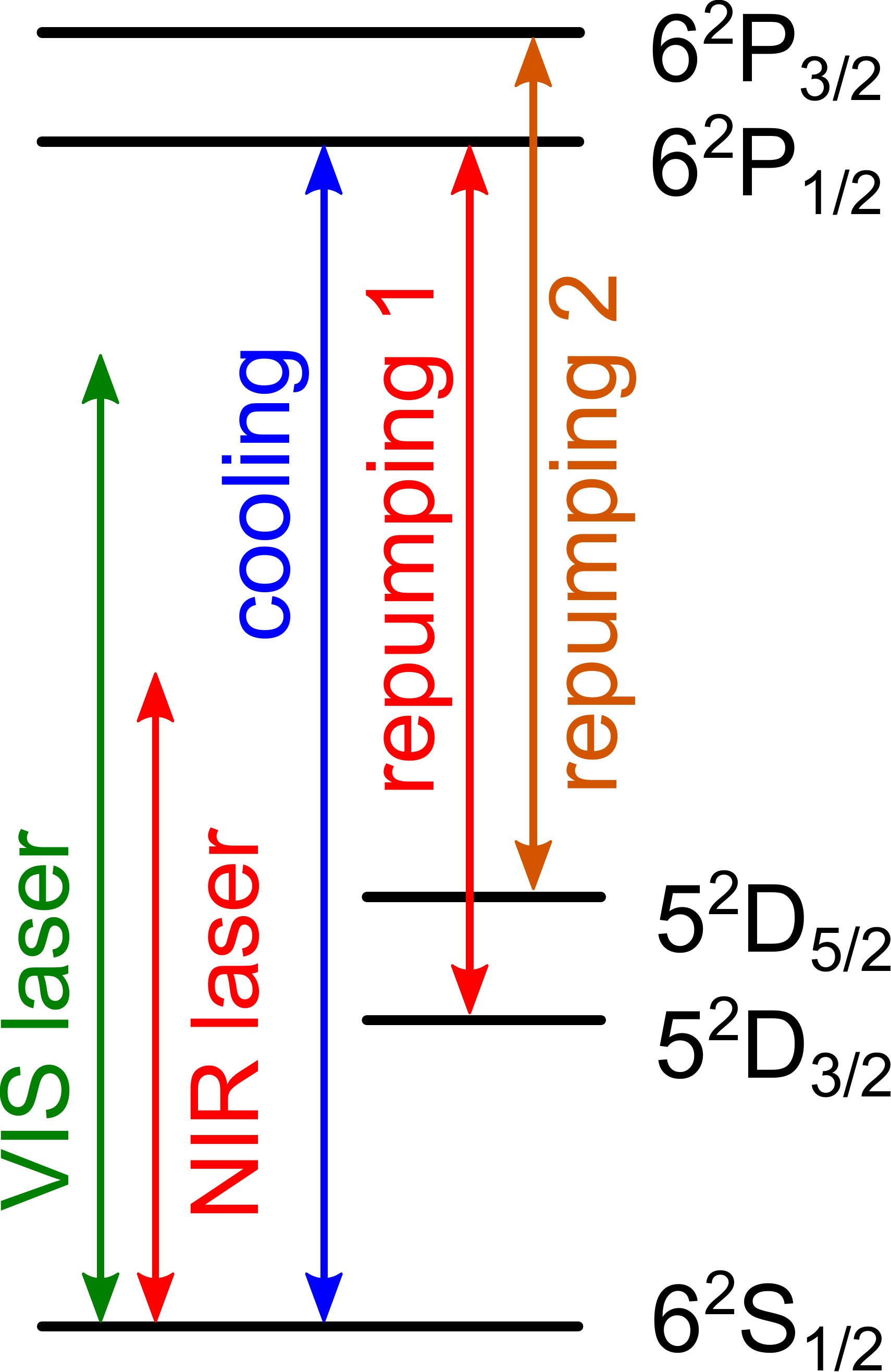}
			\label{fig:BariumLevels}}
	\caption{ 
	\textbf{(a)} Schematic set-up for optical trapping of $ ^{138}\mathrm{Ba}^+ $ in dipole traps operated at $ \lambda_{VIS} = 532 \, \mathrm{nm} $ (VIS) and $ \lambda_{NIR} = 1064 \, \mathrm{nm} $ (NIR). In addition to lasers used for Doppler cooling (blue arrow) the ion was exposed to repumping lasers (short orange arrow) for depopulating metastable $ D $ level manifolds during the trapping duration.
	 \textbf{(b)} Electronic levels and optical transitions used for Doppler cooling $ ( \lambda_{Doppler} = 493 \, \mathrm{nm} )$, repumping $  ( \lambda_{rp1} = 650 \, \mathrm{nm}, \,\, \lambda_{rp2} = 615 \, \mathrm{nm} ) $ and optical trapping. Adapted from \cite{Lambrecht2017}.}
	\label{fig:BariumLifetime}
\end{figure}

Following this approach, single $ \mathrm{Ba}^+ $ ions were optically trapped in a far-off-resonance optical trap (FORT) at a wavelength of $ 532 \, \mathrm{nm} $ \cite{Huber2014}. The drastically increased detuning compared to \cite{Schneider2010} resulted in a reduction of the off-resonant scattering rate by three orders of magnitude and the corresponding recoil heating rate by four orders of magnitude. In comparison to the previous proof-of-principle experiments carried out with $ \mathrm{Mg}^+ $, the barium ions exhibit a richer energy level structure. In particular, the presence of metastable $5 \, D_{3/2} $ and $5 \, D_{5/2} $ levels as illustrated in figure \ref{fig:BariumLevels} allows to implement state-selective potentials. In the specific configuration used in the first experiments, all ions in the said $ D $ levels experienced a repulsive potential leading to ion loss as a consequence of the dipole trap being blue-detuned to the $ 5\, D_{3/2} \longrightarrow 6 \,  P_{1/2} $ and $ 5\, D_{5/2} \longrightarrow 6 \, P_{3/2} $ transitions. The timescale for the population of these repulsively interacting states was given by the off-resonant scattering rate, resulting in lifetimes on the order of a few $ \mathrm{ms} $. Together with the improved accuracy in the detection of stray electric fields, this performance was already of interest for future investigations in the field of ion-atom interaction in a rf-free environment.\\

However, with the potential complications arising in the regime of low temperatures and high atomic densities for certain combinations of ions and atoms, such as  $^{138} \mathrm{Ba}^+ $ and $ ^{87} \mathrm{Rb} $, an increased lifetime can afford additional flexibility in the choice of initial parameters. For example, the increased occurrence of three-body-recombination events in presence of an ion impurity \cite{Haerter2012}, can make it advantageous to operate in a regime of low density below $ 10^{12} \, \mathrm{cm}^{-3} $, where the strong reduction of elastic collision processes requires a longer interaction duration. An interesting question in view of such research perspectives is what lifetimes realistically can be achieved. A series of experiments addressing this issue has recently been carried out using an experimental setup suitable for such investigations, as depicted in figure \ref{fig:LifetimeSetup}.\\

\newpage
As outlined in \cref{chpt:ODTs}, \cref*{sec:Methods} an increased performance of optical traps can be achieved by using a Paul trap with high numerical aperture access to the ion as well as by improving the accuracy of stray electric field compensation and cooling during the preparation phase. Focusing dipole traps to smaller radii increases the available restoring force for a given intensity, enhancing the effective trap depth. According to equation \ref{eqn:radialcutoff} the trapping probability scales exponentially with the ratio $ \xi^{-1} = k_B T / U_0  $ such that a lower initial temperature leads to a significantly higher trapping probability for a given dipole trap depth, as illustrated in figure \ref{fig:BoltzmannT}. The combination of these techniques allows to use comparatively low intensities at a constant level of performance, effectively reducing the off-resonant scattering rate and hence ion loss due to population of the $ D $ states.\\

\begin{figure}[h!]
	\subfigure[]{
	\includegraphics[height=0.4\textwidth]{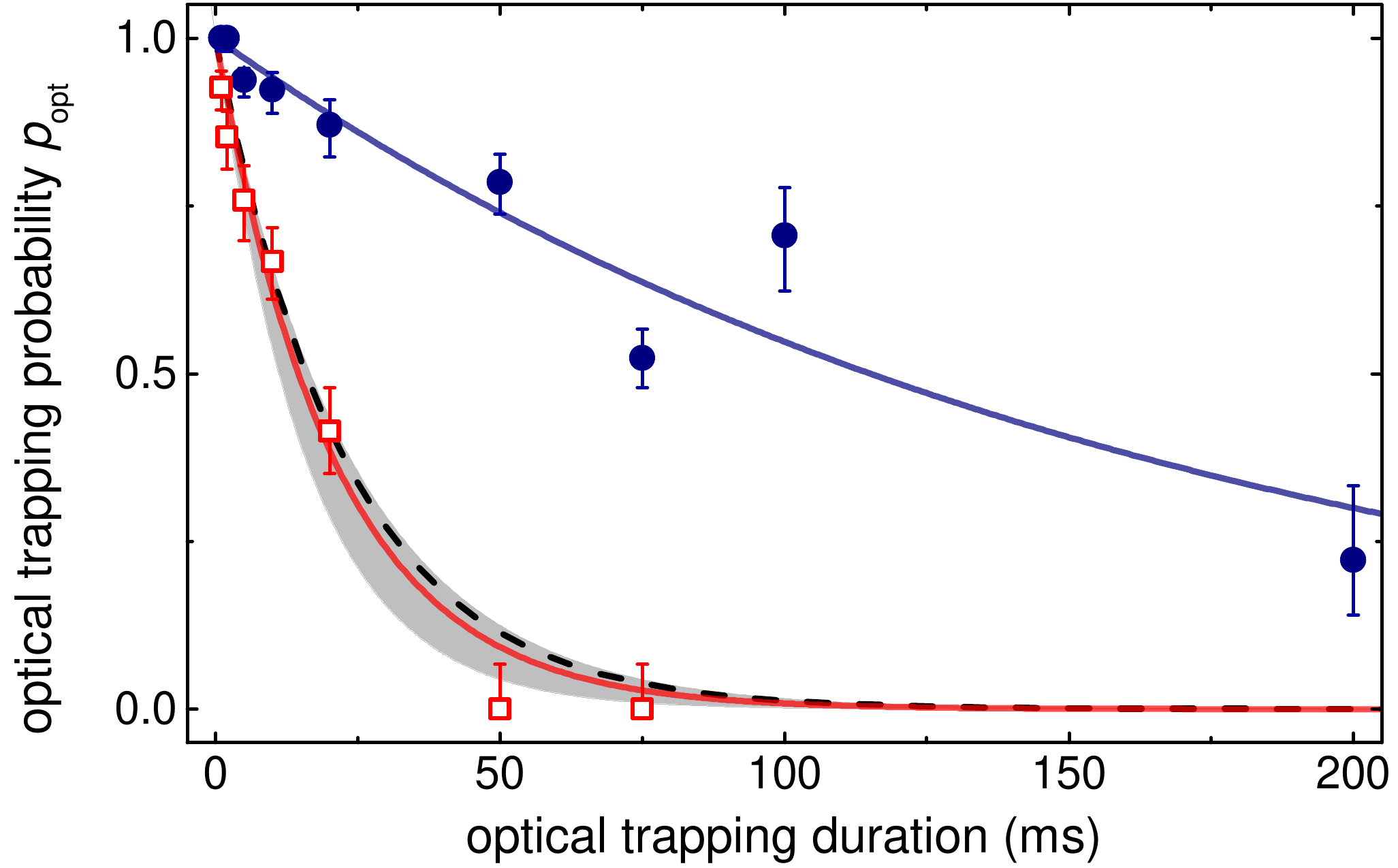}
			}
	\subfigure[]{
	\includegraphics[height=0.4\textwidth]{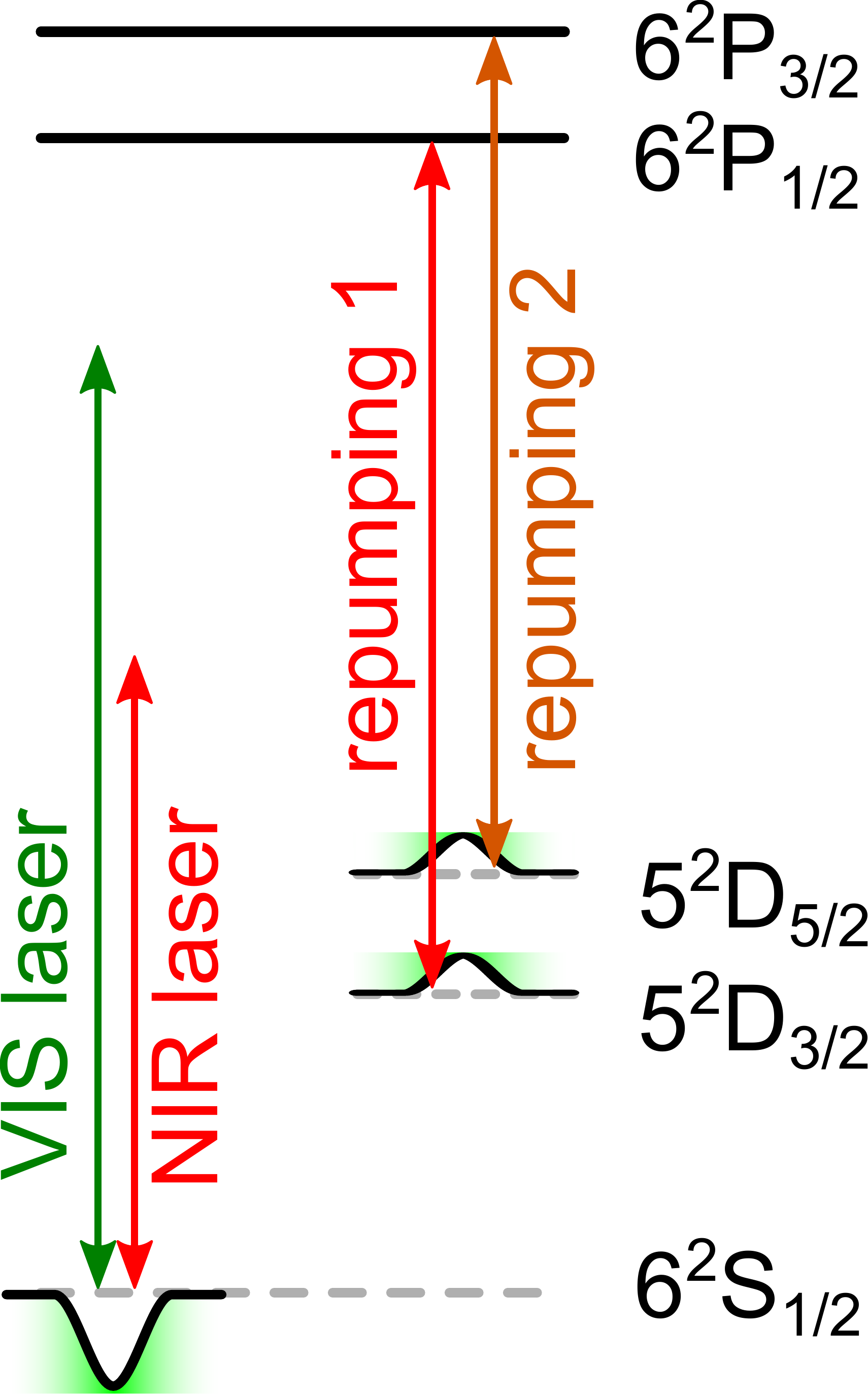}
			}
\caption{ \textbf{(a)} Dependence of $ p_{opt} $ on trapping duration $ \Delta t_{opt} $. Open squares: data taken with the dipole laser only. Dashed line: expected lifetime of $ \tau_{D} = 23 \pm 4 $ ms based on measured off-resonant scattering rates into the $ D $ manifolds. Shaded area: bounds of a theoretical prediction $ \tau_{th} = 20 \pm 4 $ ms. Circles: data taken with additional repumping lasers. Solid lines: exponential fits to each data set, showing an increase of the lifetime in the dipole trap at 532 nm,  $ \tau_{VIS} = 21 \pm 2 \, \mathrm{ms} $ to  $ \tau_{VIS, rp} = 166 \pm 19 \, \mathrm{ms} $ due to repumping. \textbf{(b)} Schematic of the relevant electronic states coupled by the dipole traps and the repumping lasers. The light shifts induced by the VIS dipole trap in the $ S $ state (lowering the unperturbed energy) and in the $ D $ states (increasing the energy). Adapted from \cite{Lambrecht2017}. }
\label{fig:VISlifetime}
\end{figure}

Implementing these improvements as compared to the first experiment demonstrating a FORT for $ \mathrm{Ba}^+ $ ions \cite{Huber2014}, in particular reducing the beam waist radius from $ 3.9 \, \mu  \mathrm{m}$ to $ 2.6 \, \mu \mathrm{m}$ and lowering the temperature after laser cooling from $ 8.5 \, \mathrm{mK} $ to the Doppler limit of $ 365 \, \mu \mathrm{K} $, the $ 1/e $ lifetime of an ion in an optical trap was significantly extended to $ \sim 20 \, \mathrm{ms} $, as shown in figure \ref{fig:VISlifetime}. It has been shown that population in the $ D $ levels originating from off-resonant scattering can be transferred back to the initial $5 ~ S_{1/2} $ state by illuminating the ion with additional repumping lasers during the optical trapping phase. Consequently, the corresponding reduction of ion loss is reflected in a sizeable enhancement of the lifetime by about an order of magnitude, in comparison to the case where no repumping is applied, to about $ 170 \, \mu \mathrm{s} $. It should be noted that the branching ratio of the $ P \longrightarrow S $ and the $ P \longrightarrow D $ transitions is about $ 3:1 $, which is quite unfavourable in comparison with other commonly employed species such as $ \mathrm{Yb}^+, \, \mathrm{Sr}^+, \, \mathrm{Ca}^+ $. The same experiment carried out e.g. with $^{171} \mathrm{Yb}^+ $ featuring a branching ratio from $ 2 ~P_{1/2} $ to $ 2~ D_{3/2} $ state of $ \sim 0.005 $, is expected to show a suppression of the loss probability via the metastable state by a factor of approximately $ 67 $ per scattered photon. Elements such as  $ \mathrm{Be}^+, \mathrm{Mg}^+, \mathrm{Zn}^+, \mathrm{Cd}^+$, and $ \mathrm{Hg}^+$ provide a closed cycling transition, such that off-resonant scattering can lead to recoil heating but not to the population of repulsively interacting electronic states and thus to loss.\\

In view of potential applications in the realm of quantum information processing or quantum simulations, it is important to note that the coherence time in the dipole trap $ \tau_{coh} $ is not prolonged by repumping. In fact, any coherent evolution is expected to be disrupted with the occurrence of the first off-resonant scattering event. A viable strategy for increasing $ \tau_{coh} $ would be to lower the initial temperature by implementing standard ground state cooling techniques, which in principle are compatible with optical potentials \cite{Kaufman2012,Karpa2013}. Making use of blue detuned standing waves may also significantly reduce off-resonant scattering by confining the ions in the nodes of the optical field \cite{Karpa2013}. An alternative route would be to implement even further detuned dipole traps, or ultimately quasi-electrostatic traps (QUEST) with scattering rates below $ 10^{-3} \, \mathrm{s}^{-1} $, for example by using $ \mathrm{CO}_2 $ laser sources \cite{Grimm2000}.

\section{Lifetime of an ion in a further detuned optical trap}
As shown in the previous chapter, the main limitation with respect to lifetimes of $ \mathrm{Ba}^+ $ ions in optical traps is related to the off-resonant scattering rate $ \Gamma_{offr} $. An effective and straightforward approach to reducing $ \Gamma_{offr} $ has been laid out decades ago in experiments with neutral atoms. As implied by equations \ref{eqn:Uopt} and \ref{eqn:GammaOpt}, it is advantageous to increase the detuning of the dipole trap, while increasing its intensity in order to maintain a constant trap depth. In the described experiment, this effect can be directly observed under identical conditions in the same apparatus by comparing the lifetime observed in a VIS dipole trap at $ 532 \, \mathrm{nm} $ with results obtained in a NIR dipole trap at $ 1064 \, \mathrm{nm} $.\\
\begin{figure}[h]
	\begin{center}
		\includegraphics[height=0.4\textwidth]{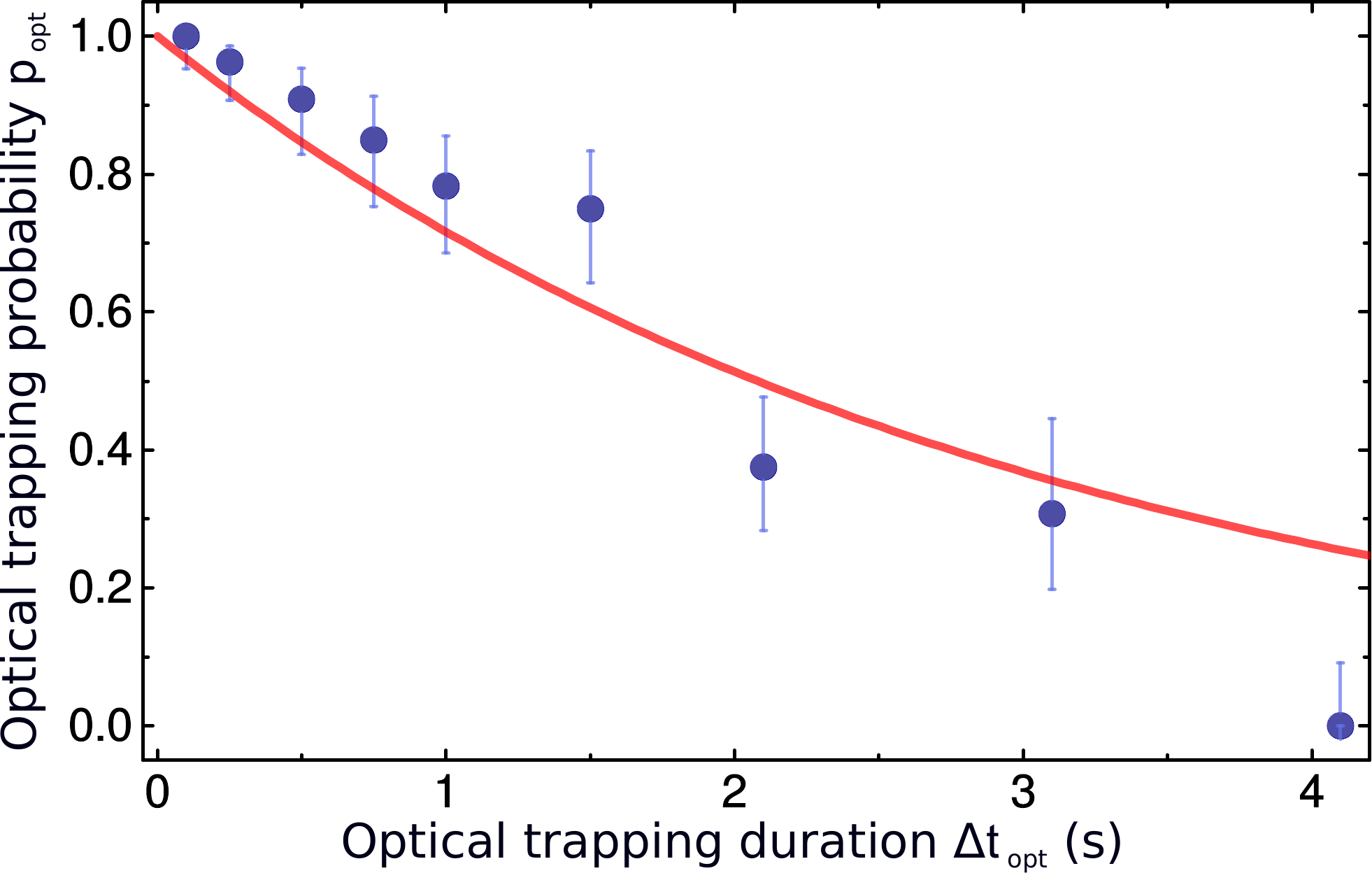}
	\end{center}
	\caption{ion lifetime in the NIR trap. Dependence of optical trapping probability $ p_{opt} $ on trapping duration $ \Delta t_{opt} $ for single $ \mathrm{Ba}^+ $ ions. Line: result of a fit, assuming exponential decay, yielding a lifetime of $ \tau_{NIR} = (3 \pm 0.3) \, \mathrm{s} $. Taken from \cite{Lambrecht2017}.}
	\label{fig:NIRlifetime}
\end{figure}

As shown in figure \ref{fig:NIRlifetime}, the change in detuning from $ \Delta_{VIS} = 3 \times 10^{6} \, \Gamma $ to $ \Delta_{NIR} = 2 \times 10^{7} \, \Gamma $ leads to a dramatic increase of the lifetime to $ \tau_{NIR} = 3 \pm 0.3 \, \mathrm{s}$, now reaching trapping durations on the order of seconds. The observed lifetimes are comparable with the results achieved in experiments carried out in a comparable regime with single neutral atoms in dipole traps \cite{Barredo2016,Kaufman2012,Endres2016,Xia2015}. This may indicate that the lifetime could be limited by collisions with background gas, rather than due to fundamental effects or other mechanisms specific to the investigated system. For example, intensity fluctuations of the dipole trap lead not only to heating as is the case in neutral atom systems, but also to a displacement of the ion. Similarly, fluctuations of ambient fields, result in random changes of the ion's position within the dipole trap, accompanied by correlated intensity fluctuations.\\

\section{Heating rate due to ambient fields}

\begin{figure}[h!]
	\begin{center}
		\includegraphics[height=0.4\textwidth]{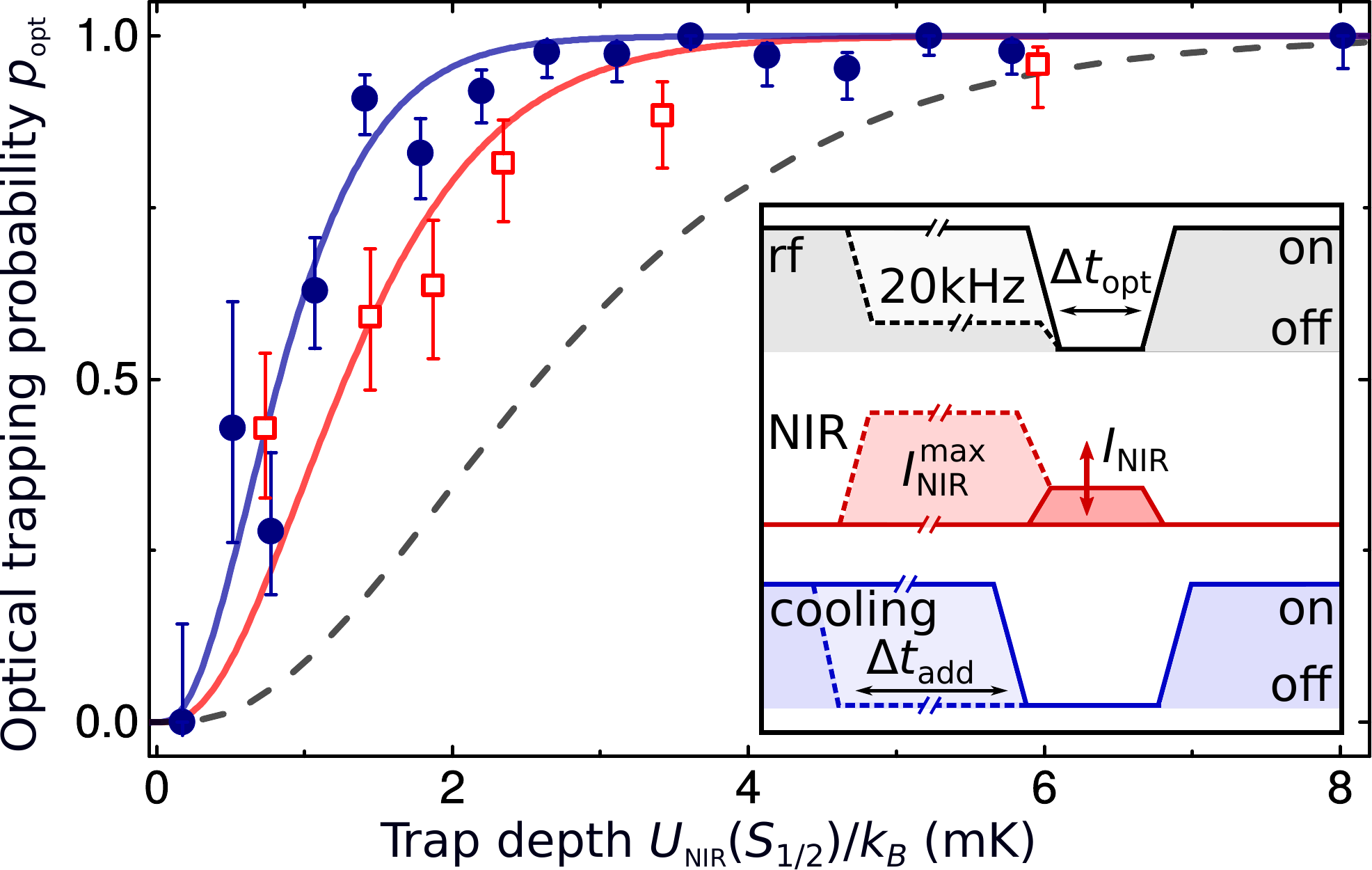}
	\end{center}
	\caption{	
	Heating rate measurement in the dipole trap. 
	$ p_{opt} $ is measured for optical trapping directly after Doppler cooling of the ion (blue circles) and after a delay of $ \Delta t_{add} = 500 \, \mathrm{ms} $, including exposure of the ion	to an optical field with maximal intensity $ I_{NIR,max}  $, ambient and weak radiofrequency fields (red squares). During $ \Delta t_{add} $ laser cooling was turned off. The temperature was probed for $ \Delta t _{opt} = 10 \, \mathrm{ms} $ by following the sequence depicted in the inset. Solid lines: fits to the data assuming a radial cutoff model, with temperatures $ T_1 = 320 \pm 30 \, \mu \mathrm{K} $ and $ T_2 = 500 \pm 60 \, \mu \mathrm{K} $. Taken from \cite{Lambrecht2017}.}
	\label{fig:HeatingRate}
\end{figure}

In order to investigate the influence of the environment including the coupling of the trapped atom to electric fields, the optical trapping sequence was adapted by including an additional waiting time of $ 500 \, \mathrm{ms} $, allowing to use the temperature measurement method described in section \ref{sec:Methods} to determine a heating rate. During this additional period, the ion was exposed to the maximum available intensity of the dipole trap, as well as to a residual rf field as indicated in the inset of figure \ref{fig:HeatingRate}. The result of such a measurement shown in figure \ref{fig:HeatingRate} therefore provides an upper bound for the heating rate of $R_{max} = 400 \, \mu \mathrm{K} \, \mathrm{s}^{-1} $. Under realistic conditions, even at the Doppler limit much lower intensities can be used to confine the ion, and can be reduced further by implementing sub-Doppler cooling techniques. In addition, a substantial reduction of the heating rate can be expected by implementing the following optimizations not adapted in the conceptual investigations described above: stabilization of laser intensity and beam pointing, optimization of electric noise filtering, and improved shielding from ambient fields by adapting the trap geometry.\\

It should also be noted that in envisioned ion-atom collision experiments, which is one of the highlighted scenarios that could strongly benefit from absence of rf fields and the related micromotion-induced heating, elastic collisions are expected to be a highly efficient cooling mechanism \cite{Tomza2015,Krych2011} with cooling rates that are several orders of magnitude higher than $ R_{max} $. For the case of $ \mathrm{Ba}^+ $ ions immersed in a cloud of ultracold $ \mathrm{Rb} $ atoms, the sympathetic cooling rate is estimated to be on the order of $ 100 \, \mathrm{mK} \, \mathrm{s}^{-1} $ \cite{Lambrecht2017}.\\

In view of potential applications in the field of quantum simulations and information processing it is instructive to compare these findings with typical heating rates obtained in planar traps as one of the most promising currently available platforms in this field of research. While it is known that proximity to electrode surfaces can result in the so-called ``anomalous heating'' \cite{Turchette2000,Deslauriers2006}, recent developments show that this effect can be reduced by one or more than two orders of magnitude by cooling the electrodes \cite{Deslauriers2006} or by employing techniques such as argon-ion-beam bombardment \cite{Hite2012}, respectively. For example, it was shown that the heating rate for a $ \mathrm{Be}^+ $ ion positioned $ 40 \, \mu \mathrm{m} $ above the chip surface can be lowered from initially $ \sim 7000 $ quanta $\mathrm{s}^{-1} $ to $ \sim 40 $ quanta $\mathrm{s}^{-1} $ for a secular frequency of $ 2 \pi \times 3.6 \, \mathrm{MHz} $ or about $ 7 \, \mathrm{mK} \, \mathrm{s}^{-1} $ \cite{Hite2012}. In comparison, the upper bound for the heating rate measured in the optical trap is substantially lower, such that optical traps for ions may be a promising complementary approach to other currently pursued techniques. However, the coupling of electric field noise to ions confined in an optical potential is expected to be non-trivial and at this point it is not clear what effects can be expected when optical traps are operated close to or in combination with electrode structures. Therefore, further experimental investigations of decoherence and heating mechanisms in optical dipole traps for ions are indispensable in order to quantitatively gauge the potential gain by applying these novel methods to QIP and quantum simulations.
\chapter{Optical trapping of Coulomb crystals} \label{chpt:CrystalsODTs}

So far, optical traps in the previously described experiments have been used to trap a single ion. As was shown in the previous chapters, the techniques adapted from neutral atom traps to ions are straightforwardly applicable, if the additional contributions to the effective potential originating from electrostatic defocusing (see figure 
\ref{fig:PotStatOptDeconf}) and stray electric fields are properly taken into account. Since no fundamental limitations with respect to the achievable performance have been identified, an interesting question concerning the scalability of optical traps to larger numbers of ions arises. This is a particularly topical question given current efforts invested in scaling the capacities of platforms for quantum information processing based on ions,  especially in light of the arising prospects of trapping two and three dimensional ion crystals without the detrimental effects of micromotion.\\

\begin{figure}[h!]
\begin{center}
	  \includegraphics[height=0.4\textwidth]{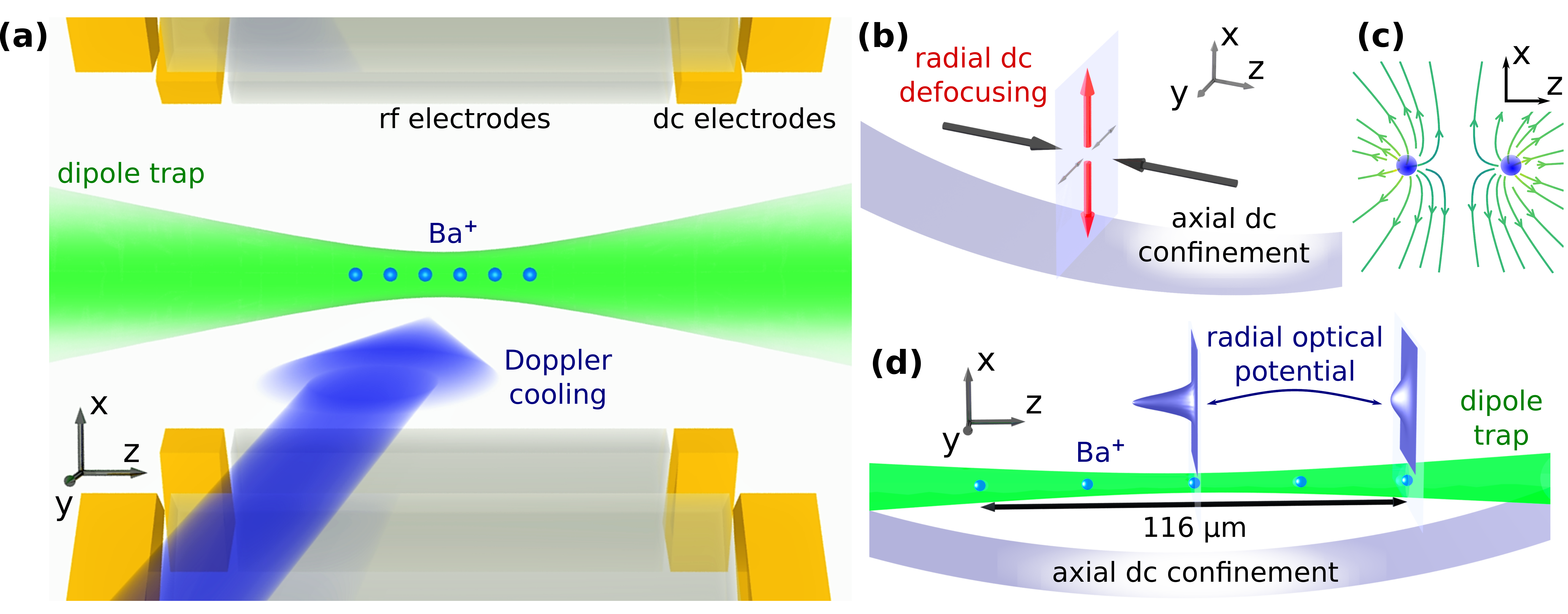}
\end{center}
          \caption{Schematic of the experimental set-up used for optical trapping of Coulomb crystals.
          \textbf{(a)} In the experiments, crystals containing up to six $ \mathrm{Ba}^+ $ ions, with lengths of up to $135\,\mu \mathrm{m}$ (5 ions: $116\, \mu \mathrm{m}$) were prepared and optically trapped. 
          \textbf{(b)} Electrostatic contributions to the effective potential due to axial dc confinement (grey surface). The defocusing mainly affected the $x$ direction (red arrow), and was negligible in the $y$ direction (thin gray arrow). \textbf{(c)} Additional defocusing in the radial directions due to ions' mutual Coulomb interaction.
          \textbf{(d)} The dipole traps were produced by focused Gauss beams (Rayleigh lengths $z_R^{VIS} = 40\,\mu \mathrm{m}$ and $z_R^{NIR} = 74\,\mu \mathrm{m}$). Therefore, the radial optical potential (blue surfaces) for the outer ions was significantly lower and shallower than for the center ion. Taken from \cite{Schmidt2018}.}
 \label{fig:CrystalTrapSchematic}
\end{figure}

As will be discussed in the following, recent experiments have demonstrated that even comparatively simple experimental arrangements based on focused beam dipole traps can be used to trap linear chains of ions as depicted in figure \ref{fig:CrystalTrapSchematic}(a). A new major aspect that has to be considered as compared to traps for single ions, is the mutual Coulomb repulsion of the constituents leading to additional electrostatic defocussing as illustrated in figure \ref{fig:CrystalTrapSchematic}(c). At the same time the presence of Coulomb interaction now provides access to the phonon modes of the trapped crystal, which is a prerequisite for applying trapped ions for quantum computation and simulation.\\

\begin{figure}[h!!]
	\begin{center}
	\includegraphics[width = 0.6 \textwidth]{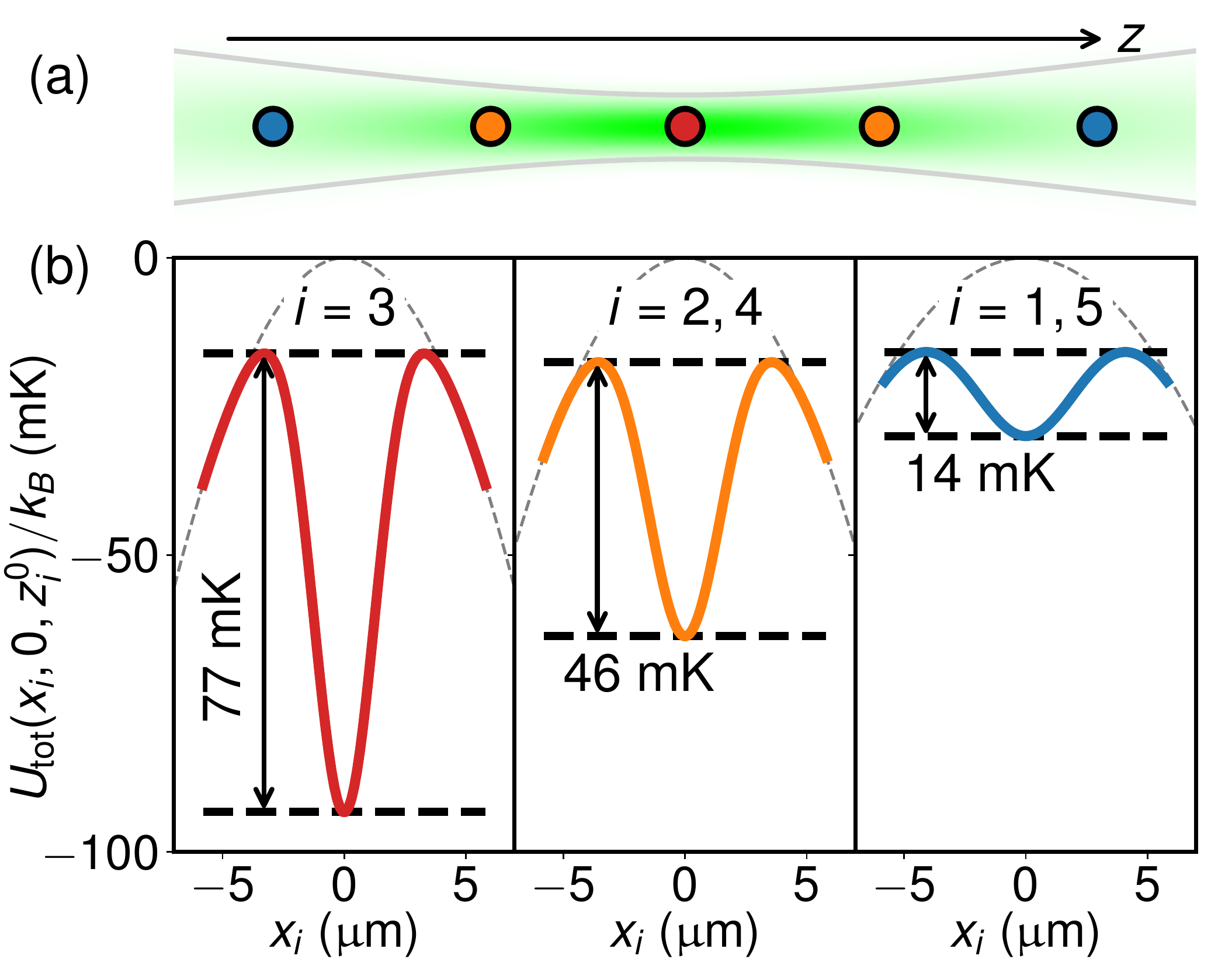}
	\end{center}
 	\caption{\textbf{(a)} Illustration of a five ion crystal and \textbf{(b)} the effective potentials experienced by ions at positions $z_i^0$.
 The ions denoted with indices $i=1...5$ (blue, orange and red circles) are exposed to different optical potentials (assuming a Gaussian dipole trapping beam with power $P^{VIS}_{opt} = 8 \, \mathrm{W}$ and waist radius $w^{VIS}_x(z=0) = 2.6 \, \mu \mathrm{m}$) depending on their axial position along $z$.
 The electrostatic potential $U_{el}(\vec{r}_i)$ (dashed gray line) is approximated as a quadrupole potential and combines the contributions of an external electrostatic potential $U_{dc}(\vec{r}_i)$ (maximum defocusing along $x$-axis) and the repulsive Coulomb interaction $U_{coul}(\vec{r}_i)$ between the ions.
 The effective potential is calculated at the position of the ions with $i=3$ (red), $i=2,4$ (orange) and $i=1,5$ (blue).
 Taken from \cite{Schmidt2018}.}
\label{fig:CoulIndividualAxialPotentials}
\end{figure}

The corrections to the total trapping potential as given in section \ref{eqn:TotalPotential} mainly consist of a generalized expression for the electrostatic contribution $ U_{dc} $ at the axial center of the trap, which now depends on position $ \vec{r}_i=(x_i,y_i,z_i) $ and includes a defocussing potential stemming from the Coulomb repulsion between the ions $ U_{{coul}} $: 
\begin{equation}
U_{{el}}(\vec{r}_i) = U_{{dc}}(\vec{r}_i) + U_{{coul}}(\vec{r}_i).
\end{equation}
Consequently, the approximated curvature of the electrostatic potential $U_{{el}}(\vec{r}_i)$ reads $\tilde{\omega}_{x,{el}}^2(z^0_i)= \tilde{\omega}_{x,{dc}}^2+\tilde{\omega}_{x,{coul}}^2(z_i^0)$ for ion $i$ in the weakest confined direction $x$.
With the discussed extensions of the potential model, the total radial potential energy for ion $i$ at the axial position $z^0_i$ is then given by:
\begin{equation}
U_{{tot}}(x_i,y_i,z_i^0) = U^{{VIS}}_{{opt}}(x_i,y_i,z_i^0) + U_{{el}}(x_i,y_i,z_i^0),
\end{equation}
where the radial optical potential  $ U^{{VIS}}_{{opt}}(x_i,y_i,z_i^0) $ experienced by ion $ i $ also depends on its axial position $ z_i^0 $. This results in a trapping potential that is weakest for the outer ions as outlined in figure \ref{fig:CoulIndividualAxialPotentials} and illustrates that the expected length scale for trappable Coulomb crystals will be set by the divergence of the employed Gaussian beam, and intuitively should be on the order of the Rayleigh range $ z_R $.\\

A consequence of the position dependence of the effective potential is the increased sensitivity of the trapping performance to the alignment of the dipole trap beam with the $ z $-axis of the Paul trap and to the beam quality. The latter can significantly deviate from the ideal radial profile of a Gaussian mode away from the focus. In addition, the contributions of stray electric fields have been neglected in the description of the effective potential. Furthermore, the techniques used for stray field compensation are most effective for linear electric fields measured at the center of the crystal, whereas higher order contributions or spatial field distributions require modified methods for detection, such as the concepts proposed in the following chapter, and a suitable configuration of compensation electrodes. These complications can lead to an effective reduction of trap depth for ions close to $ \sim \pm  z_R $.\\

\begin{figure}[h!]
	\begin{center}
		\includegraphics[height=0.5\textwidth]{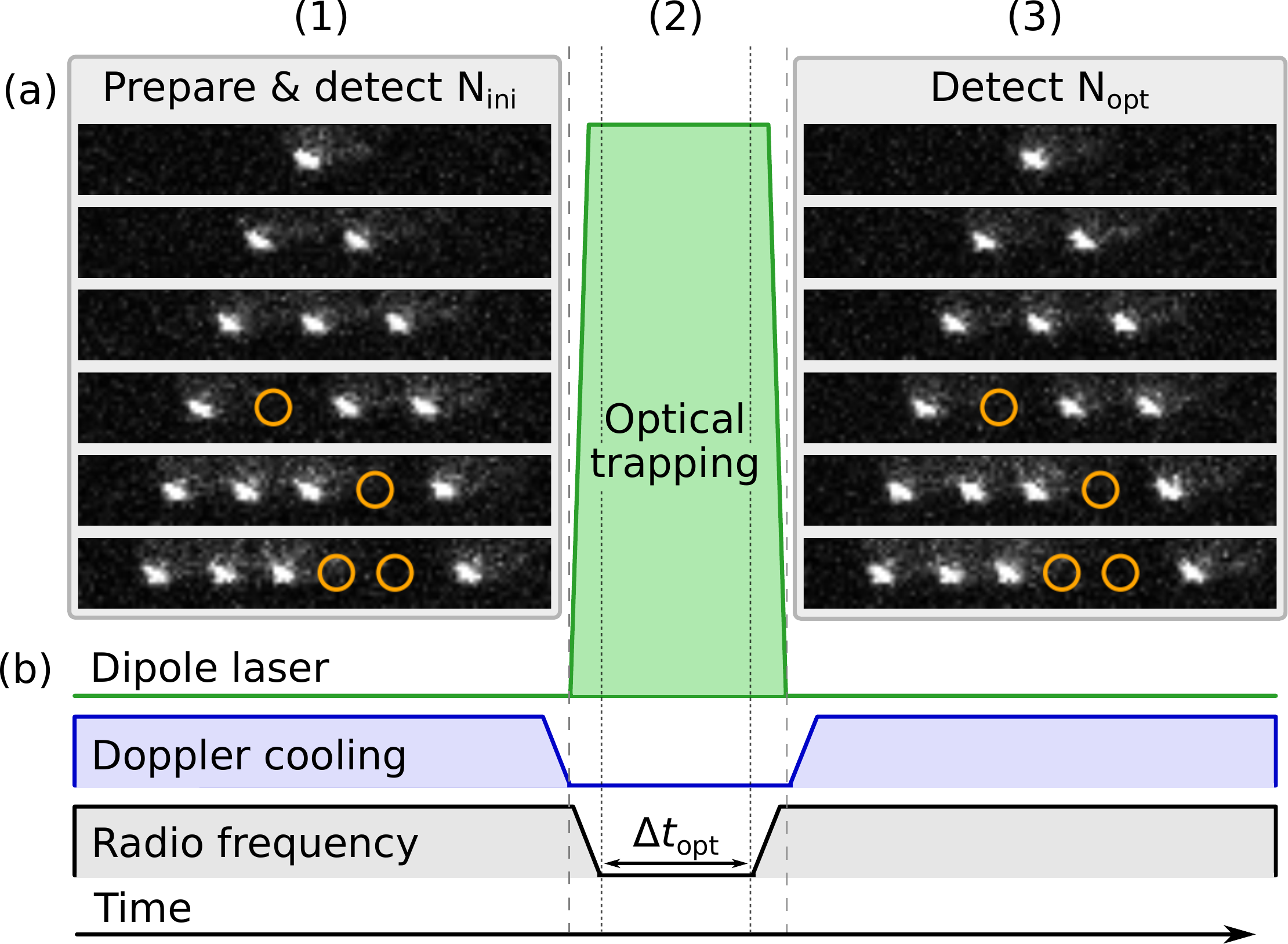}
	\end{center}
	\caption{
		Persistence of Coulomb order for an increasing number of optically trapped ions. (a) Fluorescence images of Coulomb crystals with $N_{ini} = 1...6$ $\mathrm{Ba}^+$ ions are recorded before and after optical trapping of $N_{opt}$.
		For $ N_{\mathrm{ini}} \ge 4$, the gaps marked by orange circles reveal the presence of dark ions which appear at initial random lattice sites after Doppler and sympathetic cooling.
		(b) The experimental protocol (not to scale) consists of three steps: (1) we detect the initial configuration and ion number $N_{\mathrm{ini}}$ while Doppler cooling the ions;
		(2) the ions are transferred into the dipole trap by turning off the rf field and cooling lasers for the optical trapping duration $ \Delta t_{\mathrm{opt}} $, keeping the electrostatic potential;
		(3) we again detect the number and final configuration of all remaining ions in the rf trap.
		An intermittent gaseous phase followed by recrystallization or enhanced diffusion should be observable with high fidelity via changes of the positions of the dark ions within the crystal. Taken from \cite{Schmidt2018}.}
	\label{fig:TrappedCrystals}
\end{figure}

Nonetheless, the results of recent experiments depicted in figure \ref{fig:TrappedCrystals} show that Coulomb crystals with up to 6 ions can be trapped optically using essentially the protocol developed for single ion trapping outlined in section \ref{sec:Methods}. The used potential provided a maximal trap depth of $ U_{tot} (0,0,0) \sim k_B \times 100 \, \mathrm{mK} $ generated by a focused Gaussian beam at a wavelength of $ 532 \, \mathrm{nm} $, with a power of $ 10 \, \mathrm{W} $, a waist radius of $ 2.6 \, \mu m $ and a corresponding Rayleigh range of $ 40 \, \mu \mathrm{m} $. The length of the trapped chains reaches approximately $ 135 \, \mu \mathrm{m} $, which is on the order of $ 2 z_R $ and is in agreement with the qualitative expectation based on the divergence of the Gaussian beam.

Interestingly, ions that are not addressed by laser cooling, such as barium isotopes other than $ ^{138}\mathrm{Ba}^+ $ which appeared as dark gaps in the crystal, can be cooled sympathetically and are still trappable. In this case, they can act as markers witnessing a potential change of Coulomb ordering during the optical trapping attempts, as depicted in figure \ref{fig:TrappedCrystals}. Future applications e.g. in the realm of quantum simulations with ions might build upon this property to realize optically trapped ion mixtures featuring a rich and in principle individually configurable phonon spectrum. In the investigated case, the probability for obtaining an unchanged configuration after a random redistribution of ions, i.e. due to melting and subsequent recooling, was found to be $ 9 \times 10^{-10} $ for typical experimental realizations with four ion crystals. However, this result is insufficient for concluding that the trapped chains are actually crystals, since it is unclear whether the temperature of the trapped ion structure remained below the critical temperature $ T_c \approx 50 \, \mathrm{mK} $ for the transition between a crystalline phase and a plasma.\\

\section{Temperature of ions in a dipole trap}

\begin{figure}[h!!]
	\begin{center}
		\includegraphics[width=0.6\textwidth]{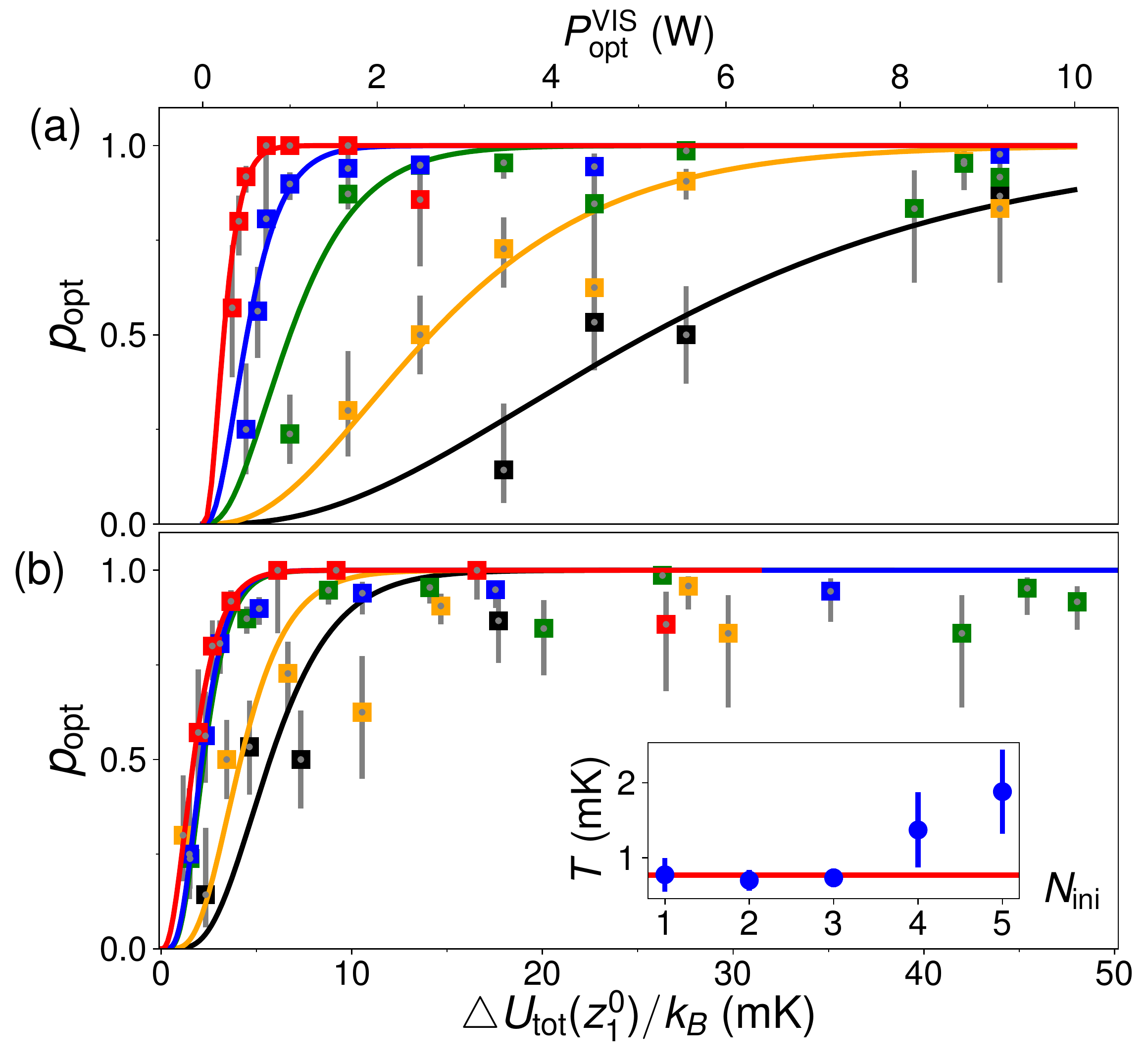}
	\end{center}
	\caption{
		\textbf{(a)} To obtain the temperature of ion chains, the trapping probability $p_{{opt}}$ was measured for $N_{{ini}} = 1...5$ ions as a function of optical power in the dipole trap $P^{{VIS}}_{{opt}}$ (red, blue, green, yellow and black squares), and the data was fitted the radial-cutoff model \cite{Schneider2012} (solid lined).
		Using the single-particle model resulted in a larger apparent temperature.
		\textbf{(b)} Analysis of the same data using an effective trap depth at locations of the outermost ions $\Delta U_{{tot}}(z^0_{1,N_{{ini}}})$, accounting for Coulomb repulsion between ions and electrostatic defocusing. Fits assuming the modified radial-cutoff model shown as solid lines, yielding temperatures of $T_{N_{{ini}}} = (0.7\pm 0.1) \, \mathrm{mK}$ $(N_{{ini}}\leq 3)$, $T_4= (1.3 \pm 0.5) \, \mathrm{mK}$ and $T_5= (1.8 \pm 0.5) \, \mathrm{mK}$ (inset). Taken from \cite{Schmidt2018}.}
	\label{fig:CrystalTemps}
\end{figure}

More stressable indications for the survival of Coulomb crystals in the dipole trap can be obtained by measuring the temperature of the sample during the optical trapping phase. The corresponding technique discussed in section \ref{sec:Methods} in principle is also applicable for more than a single ion and the results of such experiments are depicted in figure \ref{fig:CrystalTemps}. However, additional contributions from the Coulomb repulsion have to be taken into account, and the effective potential has to be calculated for each ion individually. Since $ p_{opt} $ now denotes the survival probability of the entire ion ensemble, the extracted temperatures were obtained by calculating the effective potential at the position of the outermost ion experiencing the weakest confinement. The resulting temperatures were found to be consistent with the initial values of $ \approx 1 \, \mathrm{mK} $ at the end of the preparation phase, except for the case of four and five ion chains. The apparent increase of temperature for larger ion chains could be due to the complications arising from increasing sensitivity to alignment, beam shape and stray electric fields, which are not accounted for in the model. Since all measured values were found to be at least one order of magnitude below $ T_c $, these results strongly suggest that the prepared ion chains remained crystallized during the optical trapping process.

\section{Accessing vibrational modes of Coulomb crystals in optical traps}

\begin{figure}[h!!!]
	\begin{center}
		\includegraphics[width=0.7\textwidth]{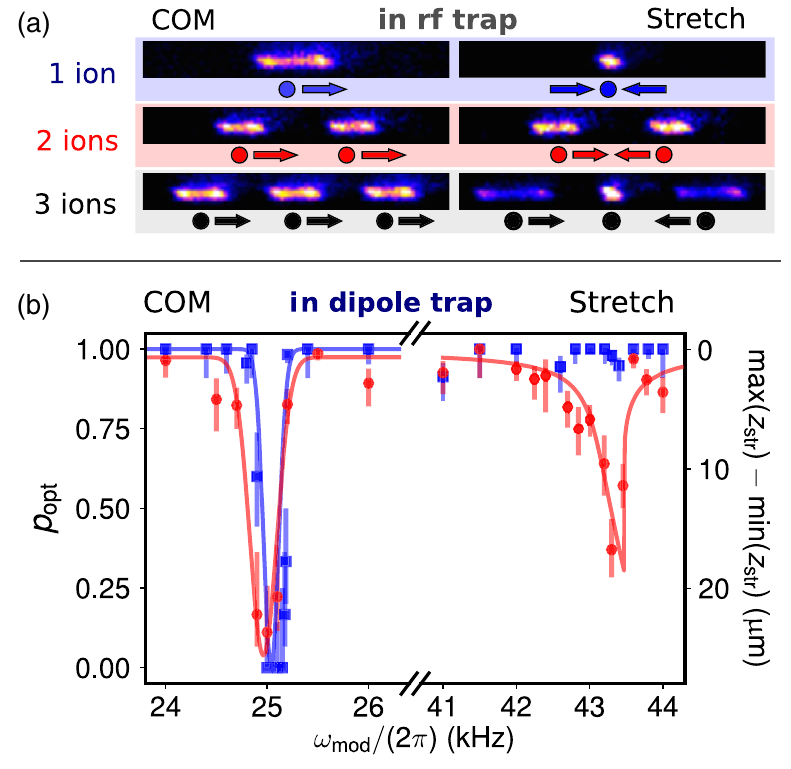}
	\end{center}
	\caption{
	  \textbf{(a)} Fluorescence images of ions in the Paul trap showing resonances excited by applying a modulation of the electrostatic axial confinement (distance between two ions is $43\,\mu \mathrm{m}$) at modulation frequencies $ \omega_{mod} = \omega_{{ax}}^{COM}$ (axial motion, $N_{{ini}}=1$), and at $\omega_{{ax}}^{{str}}= \sqrt{3}\, \omega_{{ax}}^{{COM}}$ ($N_{{ini}}=2,3$, out-of-phase motion).
	  (b) Optical trapping probability for one (blue squares) and two (red circles) ion(s)  in the dipole trap, as a function of the modulation frequency $\omega_{ {mod}}/2\pi$ of the axial electrostatic potential.
	  A drop in $p_{ {opt}}$ at $\omega_{ {ax}}^{ {COM}}$ is observed for both $N_{{ini}}=1$ and $N_{ {ini}}=2$ (solid lines: fits to data).
	  The resonance at $\omega_{ {ax}}^{ {str}} = 2\pi \times (43.3\pm0.15)\,  \mathrm{kHz}\approx \sqrt{3} \, \omega_{ {ax}}^{ {COM}}$ was observed only for $N_{ {ini}}=2$.
	  In the case of the stretch mode, we modulate the harmonic confinement to excite differential motion between the ions. 
	  The electric field amplitude at the equilibrium positions of the ions amounts to $|E|=(1.8\pm 0.1) \, \mathrm{mV}/ \mathrm{m}$ for which we numerically simulate the out-of-phase motion (no free parameters).
	  The amplitude of the stretch mode is given by $ {\mathrm{max}}(z_{ {str}}(t))- {\mathrm{min}}(z_{ {str}}(t))$ (solid red line, axis on the right-hand side). Taken from \cite{Schmidt2018}.}
	\label{fig:CrystalSpectrum}
\end{figure}
A direct signature of the crystalline phase of the trapped ions can be observed by performing an analysis of the vibrational spectrum. This can be achieved for example by periodically modulating the axial electrostatic potentials at a frequency $ \omega_{mod} $, effectively exciting the trapped ions on the corresponding normal modes \cite{Wineland2013} whenever the modulation occurs at their characteristic frequencies. For laser cooled ions confined in a Paul trap this resonant excitation manifests itself in a blurring of the ions as shown in figure \ref{fig:CrystalSpectrum}(a) for the case of the center of mass (COM) and stretch modes of up to three ions found at $ \omega_{ax}^{COM} \approx 2 \pi \times 25 \, \mathrm{kHz} $ and $ \omega_{ax}^{str} \approx 2 \pi \times 43 \, \mathrm{kHz} $ (two ion case), respectively. In order to prove the existence of these modes for optically trapped ion strings, this method has to be applied during the optical trapping phase. In a recent realization within a dipole trap operated at $ 1064 \, \mathrm{nm} $, samples of one and two ions trapped for $ \tau = 10 \, \mathrm{ms} $ were exposed to such periodic excitations. These experiments revealed sharp resonances signaled by a drop of trapping probability $ p_{opt} $ at $ \omega_{{mod}} \approx \omega_{ax}^{COM} $ for one $ (N_{ini} = 1) $ and two $ (N_{ini} = 2) $ ions, whereas an additional resonance around $ \omega_{ax}^{str} $ only could be observed in experiments carried out with two ion chains, as illustrated in figure \ref{fig:CrystalSpectrum}(b).\\

These findings confirmed that the crystals prepared in the Paul trap survived the transfer into the dipole trap, allowing for access and manipulation of their normal modes \cite{Schmidt2018} during the optical trapping phase.  

\section{Limitations and prospects} \label{chpt:CrystalsLimits}

While the first investigations highlighted in this chapter show that optical traps can be used as a new platform to investigate ion Coulomb crystals in absence of any radiofrequency fields, an important question in view of future applications pertains to the expected limitations with respect to trappable ion numbers and crystal configurations. To this end, all experiments carried out so far both with single ions as well as with Coulomb crystals provide no evidence for a fundamental limitation of the applied method. That is to say, that based on the experimental evidence available at the moment, ions can be trapped and manipulated using the same methods employed in neutral atom experiments when accounting for the contributions from mutual Coulomb interaction and electrostatic potentials, with comparable performance regarding trapping and coherence times. As shown in the case of linear ion crystals, the maximal number of trappable ions is consistent with physical constraints imposed by the chosen geometry of the dipole trap. Whereas focused beam dipole traps provide several advantages for trapping single ions, such as a comparatively simple and robust experimental set-up, specific applications may benefit from more adapted arrangements employing the superposition of several optical fields, which are well-understood and routinely used in neutral atom experiments \cite{Bloch2008,Morsch2006}. Nonetheless, the described apparatus could be extended and modified to trap higher-dimensional crystals with up to 10 ions, using standard techniques and commercially available laser sources.\\

Specifically, several intriguing ideas formulated in recent theoretical works in the field of structural quantum phase transitions \cite{Baltrusch2011,Baltrusch2012,Shimshoni2011,Shimshoni2011a} and quantum simulations \cite{Nath2015,Bissbort2013} may be within reach already with currently available ion numbers. Another recently emerged field of research where the described techniques may help accessing a new experimental regime is the study of ion-atom interactions at low and ultralow temperatures \cite{Haerter2014,Tomza2019}. So far, even the most promising approaches for reaching the quantum regime of interaction based on exploiting a favorable combination of single $ \mathrm{Yb}^+ $ ions confined in a conventional rf trap and ultracold $ \mathrm{Li} $ atoms \cite{Cetina2012,Fuerst2018} may be effective for a single ion immersed in the atom cloud, while their extension to linear ion chains is likely to be most challenging.
\chapter{Summary and Outlook} \label{chpt:Outlook}

Over the past years optical traps for ions have undergone a development from first conceptual realizations of single ion traps in proof-of-principle experiments to a point where they can be used as a novel platform featuring a number of unique properties including complete isolation from radiofrequency fields, nanoscale potentials and state-selectivity. On the one hand, all experiments carried out to date clearly show that the coupling of ions to electric fields makes it necessary to account for additional contributions to the trapping potentials as compared to neutral atoms, many of them leading to a reduced trap depth. Nonetheless, the techniques that have been developed and applied with great success in the case of neutral atoms, most notably the use of far-off-resonance traps, can also be applied for ions in dipole traps and gives rise to a comparable reduction of detrimental off-resonant scattering and the related enhancement of the trapping performance. This is exemplarily evident in the observed increase of ion lifetime from initially milliseconds to timescales on the order of several seconds achievable today. On the other hand, facing the additional challenges in comparison to neutral atom traps has also led to the development and improvement of numerous methods for manipulating ions. For example, the restrictions arising from the presence of stray electric fields have been tackled by improving upon known methods for their detection and compensation to levels below $ | \delta E_S | \approx 10  \, \mathrm{mV / m} $.\\

Of course, adapting optical traps to ions will not remain without additional challenges and drawbacks that have to be addressed in the future \cite{Cormick2011}. For instance, the Coulomb repulsion between ions in a crystal poses unavoidable restrictions with respect to the required optical powers. This puts a limit on the minimal distances and the related coupling rates. Similarly, the stray field compensation techniques successfully applied so far have not yet been adopted to the case of large ion strings or two and three-dimensional Coulomb crystals. Nevertheless, an important overarching result of previous work is that so far no fundamental limitations have been encountered, and that the main restrictions to the performance arise from technical or avoidable aspects, such as trap geometry or unfavourable branching ratios for specific ion species. As such, optical traps for ions hold promise of continued improvement, in particular with respect to achievable ion numbers, trapping and coherence time.\\   

Exploiting the features granted by the combination of optical traps and ions at the currently available level makes it feasible to approach a novel class of experiments with ions, Coulomb crystals and ion-atom mixtures in the fields of quantum simulations \cite{Nath2015}, quantum information processing \cite{Zoller2000}, ultracold ion-atom interactions \cite{Cetina2012} and metrology. In the following, I will highlight some novel applications that could be realized in the near future drawing on the new capabilities provided by optical trapping. 

\section{State-selective potentials}

An intriguing application of optical traps for ion crystals is the capability to directly realize state-selective potentials. In experiments with neutral atoms this is a well-established and common feature. Radiofrequency-based ion traps or Penning traps on the other hand are very difficult or potentially impossible to configure in a way that provides different confinements depending on the electronic state of individual ions. This is due to the fact that the confinement of ions in rf traps stems from the interaction of the charged nuclei with radiofrequency fields, whereas the interaction of the ions with electromagnetic fields arises from their coupling to the electrons. While rf fields enable very deep trapping potentials for ions on the order of $ k_B \times 10^4 \, \mathrm{K} $, in general these traps are independent of the ions' electronic states, although new promising approaches for the realization of state-selectivity, e.g. based on the properties of Rydberg ions \cite{Feldker2015,Higgins2017,Higgins2017a}, have been brought forward recently. At the same time the capability to exert state-selective forces is crucial for implementing quantum computation algorithms, quantum walks or quantum simulations with ions, and several methods utilizing optical forces within a common state-insensitive potential of an rf trap have been developed to this end \cite{Wineland2013,Leibfried2003,Monroe1995,Blatt2012,Friedenauer2008,Blatt2008}. As discussed in the previous chapters, optical dipole traps are intrinsically ideally suitable for this task, since in analogy to the case of neutral atoms, optical forces come about from light shifts of the electronic states of the ions. Provided a suitable configuration of the electronic level structure exists, this readily allows to individually and controllably manipulate the trapping potentials of individual ions within a crystal. For example, $ \mathrm{Ba}^+ $ ions feature metastable $ D $ levels located between the ground and excited states. Ions prepared in such metastable states experience a different potential depending on the wavelength of the dipole trap. In particular, an optical trap operated at a wavelength of $ 532 \, \mathrm{nm} $ induces an attractive potential in the electronic ground state, but ions in one of the $ D $ levels are repelled by the same optical field. Individual shelving of an arbitrary sub-set of ions in a Coulomb crystal can then be used to remove such ions with very high fidelity.\\

Another novel application of these properties opens up when a far-off-resonant red-detuned dipole trap is used. This is the case for a trap operated at a wavelength of $ 1064 \, \mathrm{nm} $, where both the $ S $ and the $ D $ states are trappable, but the trap depths differ by about a factor of 4. This can be used to realize entanglement between internal and external degrees of freedom as well as in experiments investigating structural quantum phase transitions, e.g. by preparing ions in a superposition of  $ S $ and $ D $ states \cite{Baltrusch2011,Baltrusch2012,Shimshoni2011,Shimshoni2011a}.\\

Similarly, the state-selectivity of optical traps is a key feature that could enable a theoretically proposed possibility of realizing fracton models by exploiting analogies between the relevant properties of fractons and elasticity theory of two-dimensional quantum crystals \cite{Pretko2018}. For example, the fractons, dipoles and gauge modes encountered in tensor gauge theory can be mapped onto disclinations, dislocations and phonons of two-dimensional crystals described by elasticity theory. Ion Coulomb crystals are potentially a highly suitable platform for observing this mapping, provided that the topological defects can be created, for example by selective removal of a row \cite{Pretko2018a}. This can be achieved by shelving $ \mathrm{Yb}^+ $ or $ \mathrm{Ba}^+ $ ions into metastable $ D $ manifolds while keeping the whole crystal in a dipole trap with a wavelength of $ 532 \, \mathrm{nm} $.\\

So far, all realizations of optical trapping without the assistance of radiofrequency fields have been carried out in macroscopically spaced three-dimensional traps. This is mainly a consequence of the large beam diameters used to obtain small beam waists of the dipole traps on the order of a few $ \mu \mathrm{m} $ which are highly beneficial due to the correspondingly large curvatures of the optical potential. However, with the rapid developments achieved in the field of planar ion traps, similar results could be obtained in the future by utilizing traps with integrated optical components capable of focusing to the $ \mu \mathrm{m} $ scale as demonstrated in recent work at the Massachusetts Institute of Technology \cite{Mehta2016}. Such a realization would allow to interface ideas behind QCCD-type trapping architectures designed for applications in QIP with dipole traps \cite{Kielpinski2002}, providing rf-free interaction regions that can be shifted away from the surface thereby mitigating or avoiding anomalous heating. The latter is considered as one of the main obstacles for planar trap miniaturization and scalability of ion-based QIP platforms. Along similar lines, optical traps could be used to measure heating rates in absence of rf fields, giving important insight into this active area of research complementary to currently employed techniques. The fundamental compatibility of optical potentials and planar ion trap geometries has already been demonstrated in experiments combining such traps with one-dimensional standing wave potentials \cite{Karpa2013,Bylinskii2015,Gangloff2015}.

\section{Higher dimensional Coulomb crystals in optical lattices} \label{chpt:highDimCrystals}

The only demonstration of optical trapping of more than one ion so far was achieved for a linear ion string within a hybrid trap where the radial confinement was still provided by rf fields \cite{Karpa2013,Bylinskii2015,Gangloff2015}, which are detrimental in applications involving ion-atom collisions \cite{Cetina2012}. More recently all-optical trapping of ion-crystals in absence of rf fields has been observed in focused beam far off-resonance traps (FORT) \cite{Schmidt2018}. Scaling of the system to 2d and 3d requires a means to overcome the limitations in ion numbers arising from the finite range of the axial trapping potential, as discussed in chapter \ref{chpt:CrystalsLimits}. This could be achieved in the near future by overlapping two or more optical fields to form a standing wave, as outlined in the following.\\

The most simple implementation would be based on an adoption of techniques demonstrated in \cite{Enderlein2012} by utilizing the optical potential of a one-dimensional standing wave. The latter would be axially aligned with a ring-type Paul trap configured to provide \textit{radial electrostatic} and \textit{axial rf} confinement. The advantage of this configuration consists in the avoidance of loss along the radial directions. A planar Coulomb crystal can be shaped and oriented in the radial plane by choosing parameters such that the curvature of the radial electrostatic potential is substantially weaker than the axial curvature of the rf potential. The Coulomb repulsion between the trapped ions is then absorbed at the expense of strong axial defocusing. The latter can now be overcome by the much steeper gradient of the intensity achievable in a standing wave or by superimposing two optical fields with wavelengths $ \lambda $ and $ \lambda /2 $ to generate a superlattice with an axial intensity envelope compensating the electrostatic defocusing.\\

\begin{figure}[h!!!]
						\subfigure[]{
							\raisebox{0cm}{\includegraphics[width=0.4\textwidth, angle=0] {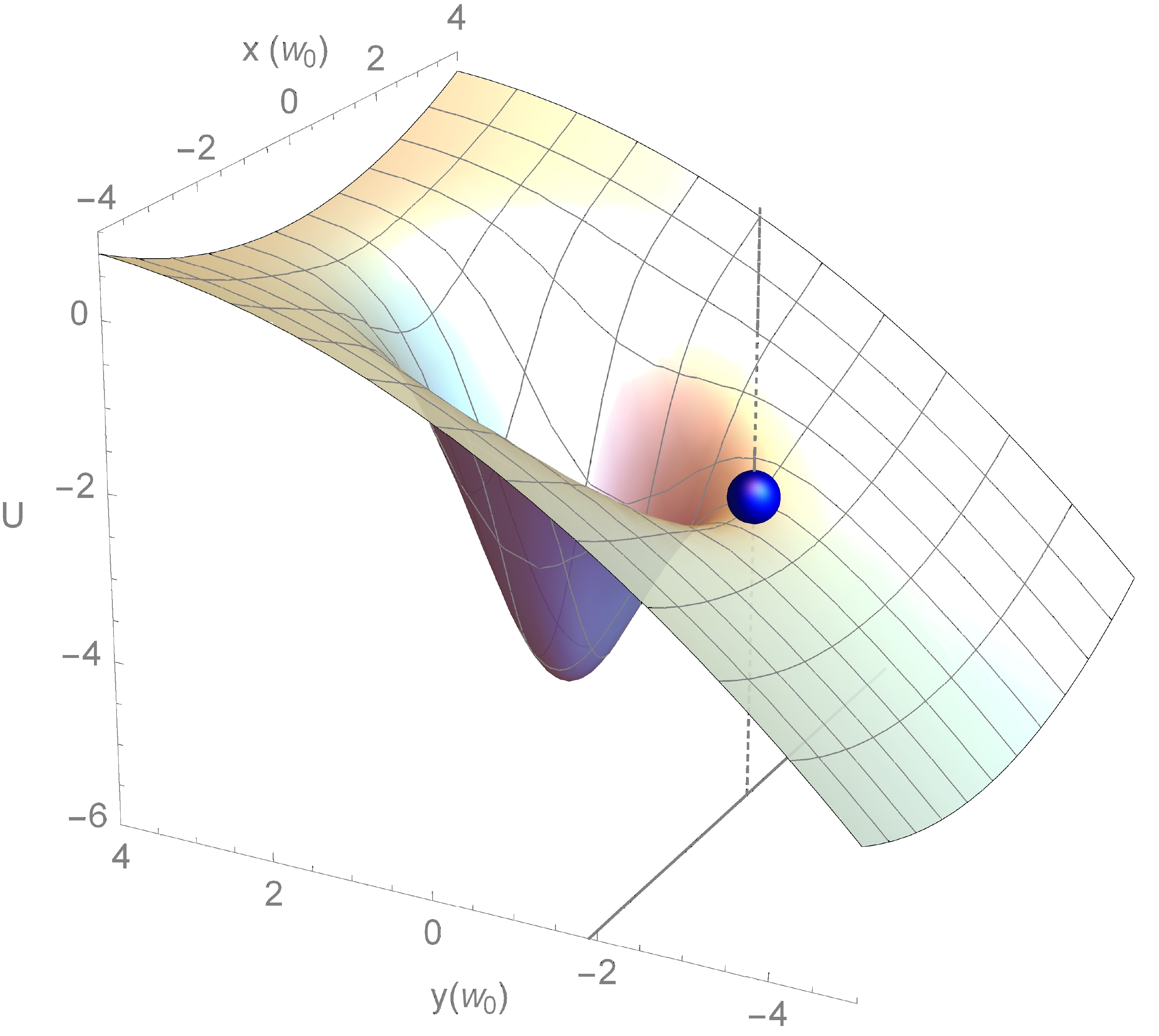}} %
							\label{fig:FocusedTrap}
						}
						\hspace{1cm}
						\subfigure[]{
							\includegraphics[width=0.4\textwidth, angle=0] {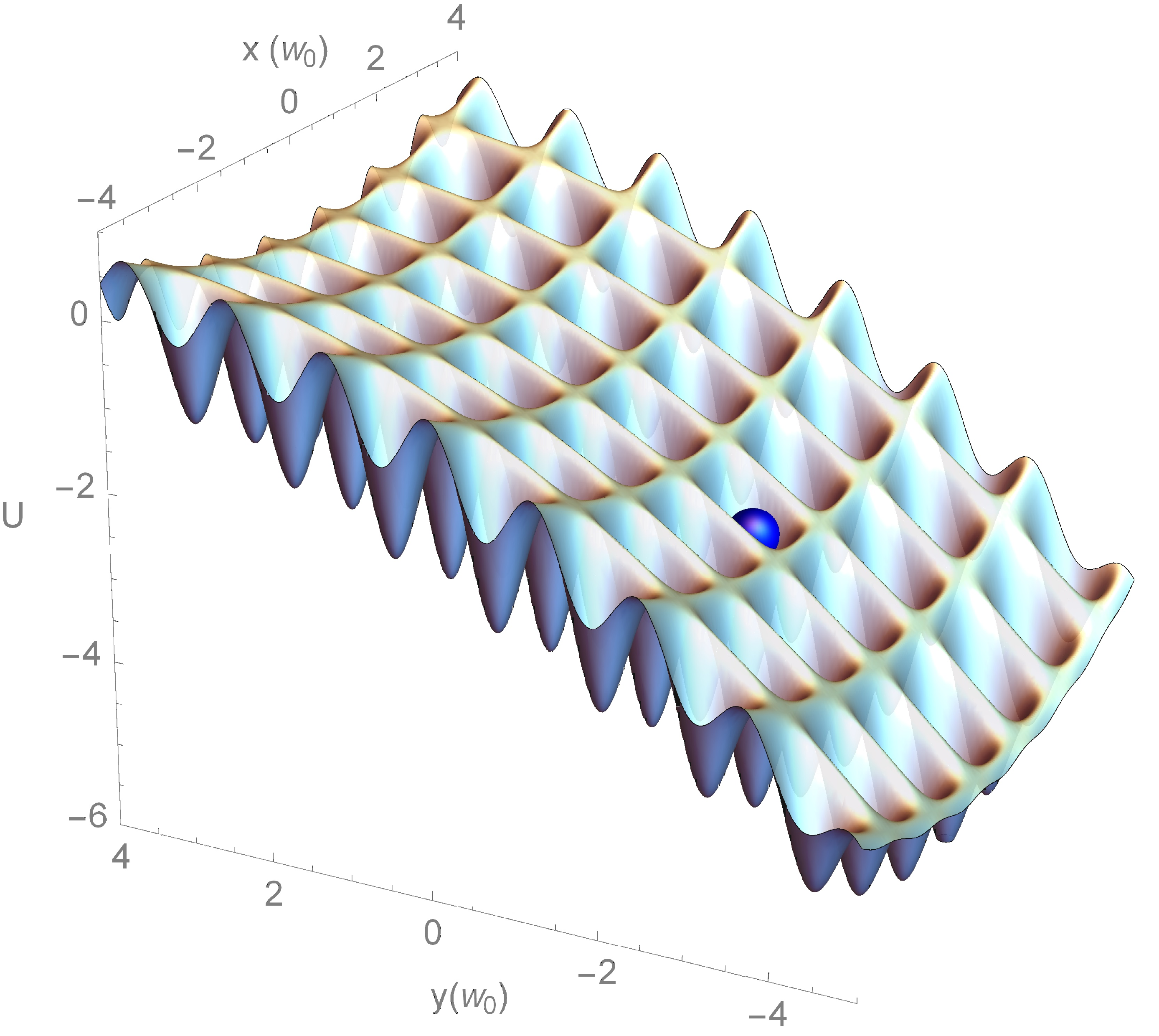}%
							\label{fig:LatticeTrap}
						}%
					\caption{\textbf{(a)} Illustration of the effective potential $ U $ of a dipole laser field focused to a waist of $w_0 $ in presence of both curvature due to the electrostatic field of the trap electrodes (additional harmonic confinement along the $ x $-axis, repulsion along $ y $) and a finite stray electric field (tilt of the potential along y), leading to reduced trap depths. An ion displaced beyond the edge of the reduced optical capture region at $y \approx -2 \, w_0 $ gets lost due to the repulsive component of the electrostatic potential. \textbf{(b)} Corresponding potential of a 2D optical lattice same electrostatic potential as in (a), obtained by superimposing two orthogonal light fields with beam waists of 5 $w_0 $ (optical power same as in (a)).
					Here, the optical forces are increased by $ \sim w_0 / (\lambda/2)$ (ratio of gradients of intensity), such that displacement due to the same stray electric field is suppressed and the ion remains trapped.					
					}
					\label{fig:DipoleTraps}
				\end{figure}

In the concrete setting described in the previous chapters, one might make use of laser sources operating at the wavelengths of 1064 nm and, after frequency doubling, 532 nm, as employed in a recent work with neutral $ ^{6}\mathrm{Li} $ atoms \cite{Kangara2018}. These two optical fields then can be arranged to produce a aligned standing waves such that at the axial center position the depth of the combined optical potential in the adjacent potential wells is initially increasing for larger distances away from the center cite. This could provide robustness against stray electric fields, which are increasingly difficult to compensate for larger crystals. For example, stray field displacement by $ 5 \, \mu \mathrm{m} $ only slightly changes the potential in a lattice site, but moves the ion out of the capture range of a focused beam dipole trap, as illustrated in figure \ref{fig:DipoleTraps}. Building upon the concepts introduced in recent theoretical proposals \cite{Baltrusch2011,Baltrusch2012,Shimshoni2011,Shimshoni2011a}, such a configuration would also be suitable for studies of structural quantum phase transitions between a 2d ion crystal and a 3d crystal by means of tunneling of the center ion, or for creating entangled states between internal degrees of freedom and structural crystal phases.\\

More complex and versatile geometries can be realized by adopting the concepts successfully applied for neutral atoms making use of several laser beams and controlling their respective degrees of freedom. For instance, superpositions of three suitably aligned beams give complex structures such Kagom\'{e} lattices. 
For neutral atoms these ideas have been extended even further, now allowing the construction of arbitrarily configurable three-dimensional micro-trap structures with more than a hundred individual sites, as demonstrated very recently \cite{Barredo2018}. In principle, all these techniques can also be applied to the case of higher dimensional Coulomb crystals. In the future, this novel experimental approach should provide a platform for investigating a to date unexplored interplay of Coulomb order and optical periodicity beyond the one-dimensional case that already has enabled to disseminate friction models in Frenkel-Kontorova-type systems at an unprecedented level of accuracy and detail \cite{Bylinskii2015,Gangloff2015}.\\

Novel and intriguing applications of the interaction between ion crystals and tailored optical potentials may also arise in the context of quantum simulations. For example, two-dimensional Coulomb crystals with a comparably small number of constituents have been recently proposed as experimental platforms for simulating spin–spin interactions on a hexagonal plaquette and quantum magnetism \cite{Nath2015}, where optical forces are used to tailor the phonon-mode spectrum. In the future, trapping of ions in designated sites of optical lattices may enable even more refined control allowing for individual shaping of the local phonon modes.

\section{Metrological applications}

\begin{figure}[b!!!]
	\begin{center}
		\includegraphics[width=0.99 \textwidth, angle=0] {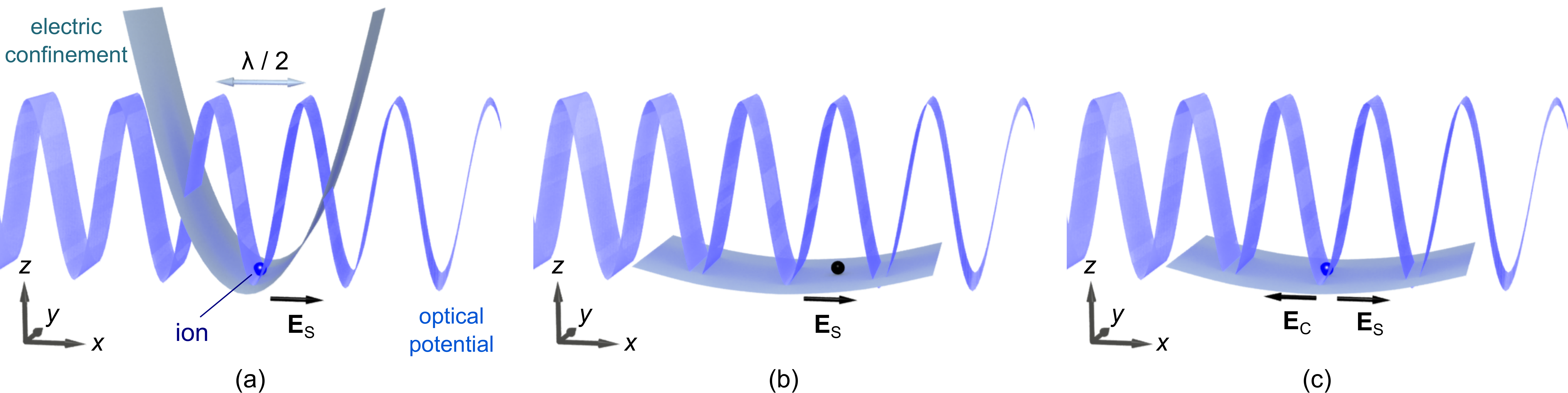}
	\end{center}
	\caption{Schematic of the procedure for compensating a stray electric field $ \textbf{E}_{S} $ based on the scheme in reference \cite{Huber2014}. \textbf{(a)} Laser cooled ions placed in the nodes of the standing wave experience virtually no ac Stark shift and scatter light at a rate $ R_0 $. Displacements from the null of the (rf) electrostatic (pseudo)potential are small due to conventional pre-compensation and the steepness of the trap. \textbf{(b)} As the electric trapping potential is lowered the ion gets displaced to positions where the ac Stark shift $ \Delta_{ac} $ is considerable, shifting it out of resonance and reducing the scattering rate to $ \sim R_0/ \Delta_{ac}^2 $. Here the sensitivity to displacements due to $ \textbf{E}_{S} $ is enhanced to a fraction of $ \lambda / 2 $ compared to the currently best compensation method with a characteristic length scale determined by the waist $ w_0 $ of the focused dipole trap beam \cite{Huber2014} ($ w_0 = 4 ~ \mu \mathrm{m} $). \textbf{(c)} Compensation is detected by measuring fluorescence while applying a compensation field $ \textbf{E}_{C} $ such that the ion is moved back to the node of the standing wave.
	}
	\label{fig:WaveComp}
\end{figure}

The methods for overlapping ions with optical fields described in the previous chapters, in particular in section \ref{sec:Methods}, may also be useful in the realm of metrological applications. For example, ion crystals aligned with standing waves could be used for sensing of external fields. To this end, in the simplest case individual positions could be mapped to spatially resolved fluorescence patterns, as illustrated for one dimension in figure \ref{fig:WaveComp}. This straight-forward method is expected to allow for detecting external field gradients and their evolution simultaneously with high spatial and temporal resolution, as determined by the spacing between ions of typically a few $ 10 \, \mu \mathrm{m} $ and the available detection sensitivity for stray electric fields of $ | \delta E_S | \approx 10 \, \mathrm{mV/m} $ after integration times of about $ 10 \, s $. In more advanced scenarios the obtained sensitivity could then be enhanced even further by employing spectroscopic methods for measuring the locally induced light shifts in the energy levels of individual ions rather then by monitoring their fluorescence. Three-dimensional maps can then be obtained by observing 3d Coulomb crystals either exposed to a 3d optical lattice or a set of orthogonal 1d standing waves switched on in sequence.\\

Another exciting approach is the combination of optical trapping techniques with surface electrode traps, either by interfacing Coulomb crystals with optical cavities \cite{Karpa2013} or by adapting recently developed methods for direct integration of optical components \cite{Mehta2016} with planar traps. These concepts could be used and extended in the near future to realize novel experimental arrangements for measuring heating rates close to electrode surfaces in absence of radiofrequency fields or, more conservatively, for creating optical patterns for sensing of electric field distributions.

\backmatter

\bibliographystyle{unsrt_LK}

\chapter{References}
\bibliography{chapter01}  

\chapter*{Acknowledgments}

The projects described in this work have received funding from the European Research Council (ERC) under the European Union’s Horizon 2020 research and innovation program (Grant No. 648330). The author is grateful for financial support from the Alexander von Humboldt Foundation and Marie Curie Actions.\\

I am indebted to all my colleagues who have contributed to several generations of optical trapping experiments. In particular, it was my privilege to work side by side with T. Huber, J. Schmidt, A. Lambrecht, and of course, T. Schätz who has conceived and initiated the experiments on optical ion trapping. I would especially like to thank Vladan Vuleti\'{c} for introducing me to the field, our numerous fruitful discussions, as well as his prudent advice and support.\\

\end{document}